\documentclass[graybox, envcountchap]{svmult}

\usepackage{mathptmx}        
\usepackage{amsmath}
\usepackage{amssymb}
\usepackage{color}
\usepackage{helvet}          
\usepackage{courier}         
\usepackage{dirtree}

\usepackage{makeidx}        
\usepackage{graphicx}        
\usepackage{subfig}

\usepackage{multicol}        
\usepackage[bottom]{footmisc}

\usepackage{hyperref}        
\hypersetup{colorlinks=true,urlcolor=blue}

\usepackage[verbose,hyperpageref]{backref}

\renewcommand*{\backref}[1]{}
\renewcommand*{\backrefalt}[4]{[{\tiny%
    \ifcase #1 Not cited.%
          \or Cited on page~#2.%
          \else Cited on pages #2.%
    \fi%
    }]}

\usepackage[misc]{ifsym}

\usepackage{pgf}
\usepackage{tikz}
\usepackage{tikz-3dplot}

\makeindex             

\begin{document}


\title{Gravitational collapse of a spherical scalar field}
\author{Roberto Giamb\`{o}}
\institute{Roberto Giamb\`{o} (\Letter) \at University of Camerino (Italy), Scuola di Scienze e Tecnologie, Mathematics Division. \email{roberto.giambo@unicam.it}
\at INAF, Sezione di Roma
\at INdAM -- GNAMPA
\at INFN, Sezione di Perugia
}
%
%
\maketitle

\abstract{
Examining the relativistic collapse of a spherical spacetime where gravity is coupled with a scalar field, this review provides a thorough analysis of some of the most relevant studies from both analytical and numerical perspectives. The discussion includes achievements made in this field, with a focus on those related to cosmic censorship, as well as recent perspectives on the topic.
}


\section{Introduction and historical context}\label{sec:begin}
From the second half of last century's 60s, the so called ''singularity theorems'' of Stephen Hawking and Roger Penrose, together with the first observational evidence of black holes candidates,  shifted attention to the formation of singularities as an outcome of a \textit{dynamical} process.  Actually, despite the official reasons that were given half a century later for awarding the Nobel Prize to Roger Penrose, it is always worth remembering that the theorems of Hawking and Penrose are rather results that assert the geodesic \textit{incompleteness} of a Lorentzian manifold. It was Hawking himself that supported (see e.g. \cite[pag. 258]{Hawking:1973uf}) the notion of causal geodesic incompleteness as a mathematically viable definition of spacetime singularity that, however, is not always completely satisfactory under the physical point of view \cite{Geroch:1968ut}. In fact, one can conceive solutions where initial data on a would--be Cauchy hypersurface evolve into an incomplete spacetime that terminates at a Cauchy horizon, i.e. a  boundary of the maximal future development -- here, solution can be possibly extended but not in an unique way. This surprising feature is observed even in seemingly simple examples as the Kerr spacetime, indicating that the formation of high curvature regions evolving into a trapped surface from regular data is far from being solved by simply resorting to geodesic incompleteness. For a comprehensive account of the history, stemming from the Penrose and Hawking theorems, let me recommend the beautiful review \cite{Landsman:2022hrn}. 

As the interest in the dynamical features of spacetimes grew, it became clear that Schwarzschild (or Kerr) spacetime could only be seen as an \textit{asymptotic} state of more realistic configurations that could potentially develop a singularity. Therefore, it was necessary to consider solutions with richer features in order to gain insight into the problem of relativistic collapse. However, the nonlinear complexity of the Einstein Field Equations suggested the need for a trade-off between mathematical and physical reasons. As a result, researchers turned their attention to the collapse of \textit{spherically symmetric} models, with a particular focus on the study of collapse in dust clouds, radiation, and scalar fields. This review will specifically examine the latter model.

Studies on scalar fields also date back to the late 1960s, with the introduction of the concept of spin-zero particles described by a scalar wave, which led to the development of the so-called \textit{boson star} -- for a comprehensive review, see \cite{Liebling:2012fv}.
From a mathematical standpoint, as we will see, the problem has been modelled in many different ways, using various gauges and assumptions. However, almost every approach shares some common features:
\begin{enumerate}
\item a metric $g$ on the 4--dimensional manifold $M=\mathcal Q\times S^2$, that is given by the product metric of a Lorentzian metric on the 2-dimensional spacetime $\mathcal Q$ and a conformal metric to the induced Riemannian metric on $S^2\hookrightarrow\mathbb R^3$, with conformal factor depending on $\mathcal Q$;
\item an energy momentum tensor arising from a Lagrangian density
\begin{equation}\label{eq:Lagr}
L=\frac12 g(\nabla\phi,\nabla\phi)+\frac12 m^2\phi^2+\widetilde V(\phi),
\end{equation}
where $\phi:\mathcal Q\to\mathbb R$ is the scalar field, $\nabla$ is the gradient induced by $g$, $m$ is the constant mass of the scalar field ($m=0$ is the so--called \textit{massless} case) and a \textit{potential} function  $\widetilde V:\mathbb R\to\mathbb R$ that can be possibly set to zero (\textit{free} case) is also introduced. 
The Lagrangian  may contain other terms when  the scalar field is coupled with other energy sources, for example an electromagnetic field (see Section \ref{sec:Maxw}).
\end{enumerate}
Therefore,  in its more general extent, in the present paper  $(M,g)$ will be such that
\begin{equation}
\label{eq:EFE}
R_{\mu\nu}-\frac12 g_{\mu\nu}R=8\pi T_{\mu\nu},
\end{equation}
holds and, using an effective potential $V(\phi)$ possibly embodying the mass contribution, \eqref{eq:Lagr} gives
\begin{equation}
\label{eq:T}
T_{\mu\nu}=\phi,_{\mu}\phi,_{\nu}-\left(\frac12 g^{\alpha\beta}\phi,_{\alpha}\phi,_{\beta}+V(\phi)\right)\,g_{\mu\nu}.
\end{equation}
It is straightforward to calculate the conservation law 
\begin{equation}
\label{eq:Bianchi}
T^{\mu}_{\nu;\mu}=\phi,{_\nu}\left(
\Box\phi-V'(\phi)\right)=0,
\end{equation}
where $\Box\phi=g^{\mu\nu}\phi_{,\mu;\nu}$ is the curved wave operator induced by $g$. The equation $\Box\phi=V'(\phi)$ is often referred to as \textit{Klein--Gordon equation}. 

To trace the beginning of research on scalar field collapse, we must start with the earlier works of Demetrios Christodoulou. As he recounts in his monograph \cite{Christodoulou:2008nj}, the problem of the evolutionary character of trapped surfaces was posed to him by his PhD mentor at Princeton, John A. Wheeler. After working on stars filled with dust matter, where gravitational attraction is counterbalanced by outward pushing forces due to internal nuclear reactions, he turned his attention to the study of the spherical scalar field. As explained in \cite{Chr99}, one reason why he chose to study this model was because the dynamics governing its evolution are precisely described by the wave equation. Therefore, one could view this model as a perturbation of some background spacetime.

Christodoulou's first work on the subject, published in \cite{Christodoulou:1986zr}, focused on the massless free case ($V=0$ in \eqref{eq:T}), as did his subsequent works. The main result of this paper is a theorem that shows that if one considers initial data that are sufficiently small, they evolve into a globally unique solution that is null geodesically complete and asymptotically equivalent to flat spacetime. This result had a significant impact on the field of research, as it showed that despite the absence of forces counterbalancing gravitational attraction, as in the case of a dust star, the collapse of a scalar field -- at least in the massless free case -- does not necessarily lead to the formation of spacetime singularities.

At the time, Christodoulou was not the only one working on scalar field collapse. Matthew W.  Choptuik had already addressed the problem using numerical methods in his PhD thesis at British Columbia, under William Unruh supervision, and  had  published a couple of papers in which  the presumed inconsistency of some numerical codes useful for numerically solving the Cauchy problem for some relativistic models was discussed and fixed. Choptuik got in touch with Christodoulou at the end of the 80s, when the latter was at Syracuse. On one hand, a picture was emerging that suggested the existence of a sort of phase transition that governed initial data of solutions, with some evolving into a singularity and others regularly dispersing in the infinite future. However, it was still unclear at the time  whether one could fine--tune the data to develop a horizon and obtain black holes of arbitrarily small mass, and whether one could fully capture the causal behaviour of  this hypothetical phase transition between black hole formation and dispersion.

Indeed,  Choptuik's numerical analysis published in \cite{Chop92,Choptuik:1992jv} confirmed the existence of what would have been subsequently known as the \textit{black hole threshold}, dividing the two classes of initial data. In other words, considering various 1--parameters family of initial data, there exists a critical value of the parameter separating the two possible evolutions -- i.e. trapped surface formation vs dispersion. This behaviour was found later to be common to many collapsing models, and for a complete account on the subject the main reference is the Living Review by Carsten Gundlach and Jos\'e M. Mart\'{i}n--Garc\'ia \cite{Gundlach:2007gc}.

The solution corresponding to the critical value of the parameter can be shown to exhibit surprising features \cite{Martin-Garcia:2003xgm}. To begin, it is a discretely self--similar solution. Moreover, it contains a point singularity that can communicate with the infinite future boundary -- it is, therefore, an example of a \textit{naked singularity}, thereby violating the cosmic censorship conjecture made by Roger Penrose \cite{Penrose:1969pc}. By the way, it was Choptuik's discovery that in 1997 put an end to the first bet on naked singularity between Hawking, who was unfavourable to their existence, and John Preskill and Kip Thorne, who supported the possibility of naked singularities. Numerical analysis were suggesting that the initial data leading to naked singularities might not be generic, and this led to a new version of the bet. Furthermore, it boosted research towards an analytical approach to the analysis of naked singularities. And this bring us back to Christodoulou's work.

After some follow-ups to \cite{Christodoulou:1986zr}, in the attempt to prove Penrose conjecture, he had published in 1991 the first of four papers that are still nowadays considered  cornerstones of the subject. The result \cite[Theorem 5.1]{Christodoulou:1991yfa} expresses sufficient condition on the initial data of the problem leading to trapped surface formation, hence to a geodesically incomplete spacetime. That is a key result for the topic under exam here and for all following research (see discussion after Theorem \ref{thm:chr1} below).
 
The solutions considered so far were sufficiently regular, i.e. at least differentiable. The second paper \cite{Chr93} of the series  improves previous results leading to dispersing behaviour but mostly takes into account bounded variation solutions, that prove to be extremely important for the other two papers of the series. Indeed, in \cite{Christodoulou:1994hg} an analytical example of a scalar field solution collapsing to a singularity without horizon formation, hence naked, is found out. In the same work it is announced, for ``a subsequent paper'', that the space of initial data leading to solutions with this feature is unstable in the larger space of bounded variation solutions considered in \cite{Chr93}. Curiously enough this last paper was published only five years later \cite{Chr99a} -- although received by the journal a couple of years before publication -- and it has since often been popularized as \textit{the proof} of Penrose conjecture. In Section \ref{sec:instability} we will get back to the influence of this work in relation to cosmic censorship. 

The methods pioneered by Christodoulou were refined for a number of more recent results, especially concerning the Einstein--Maxwell--scalar field model, see Section \ref{sec:Maxw} below.  These techniques,  sharing a common approach involving the analysis of an initial value problem 
for the PDE governing  system's evolution,  led to a number of considerable result and advances in the study of problems in Mathematical Relativity, under very general assumptions. Some literature can be traced back from \cite[footnote 297]{Lands2021}

On the other side, since the   dynamical system picture was not fully resolved by the theorems in \cite{Christodoulou:1991yfa}--\cite{Chr99a}, a significant effort was devoted to many analytical and/or numerical studies of the 'critical' case found out by Choptuik. The aim was to determine the extent to which Christodoulou's non-genericity result was essentially describing the same situation as the Choptuik black hole threshold. Despite some initial enthusiasm, it must now be concluded that the link between the two pictures, while displaying some overlapping features, has not been fully resolved to date.

On quite a different pathway, a few years later, stimulated by connections with extended gravity theories and string theory, a growing interest arose about models self interacting with a potential. Dealing with a cosmological setting, in this case one usually restricts to homogeneous models where the background metric interacting with the scalar field is a Robertson--Walker (RW) spacetime -- this gives the obvious advantage to reduce the model's equations to an ODEs system. A huge variety of potentials $V(\phi)$ has been discussed within this framework, finding conditions under which the universe collapse to a future singularity, and global models without the formation of an apparent horizon has been discussed also in this situation.  Of course, this study interacted with those aiming to find quantum mechanisms to avoid recollapsing universe, see e.g. \cite{Goswami:2005fu} -- a quite recent and complete review on this subject is \cite{Malafarina:2017csn}. 

\medskip

The review is organized as follows. The basic results from Christodoulou and PDE approach are described and commented in Section \ref{sec:PDE}. The following section \ref{sec:crit} is devoted to the advances in the study of the black hole threshold and the critical case, and the homogeneous case is described in Section \ref{sec:hsf}.
One last section \ref{sec:stuff} is dedicated to some hints  and studies that are not completely related to the topics covered in the previous section, and to sketch conclusions.

\section{Scalar field collapse as a PDE initial value problem}\label{sec:PDE}
In this section we review among the most influential and important results regarding the scalar field collapse, when the evolutionary PDE approach is embraced. As said in the Introduction, this field can be traced back to the earlier works of Christodoulou on the subject, but it must be also said that these were pioneering works, with a notation often invented specifically to solve the problem at hand, and therefore not yet consolidated in the literature at the time and  sometimes not fully optimized, so the less accustomed reader may find it challenging at first glance. For example,  in \cite[p. 345]{Christodoulou:1991yfa} a geometric Bondi coordinate system is used to describe the metric, 
\begin{equation}
\label{eq:Bondi1}
\mathrm ds^2=-a(u,s)\,\mathrm du^2-2\,\mathrm du\,\mathrm ds+r^2(u,s)\,\gamma_{\mathbb S^2}
\end{equation}
(we refer to $\gamma_{\mathbb S^2}$ as the Euclidean metric induced by $\mathbb S^2\hookrightarrow\mathbb R^3$) but many of the calculations in the same work are done in a double null frame. The use of double null coordinates is rather effective in this PDE approach:
\begin{equation}
\label{eq:dnull}
\mathrm ds^2=-\Omega^2(u,v)\,\mathrm du\,\mathrm dv+r^2(u,v)\,\gamma_{\mathbb S^2}. 
\end{equation}
Coherently with \cite{Chr93} and other works in this thread, we normalize the scalar field $\phi\to\frac1{\sqrt{4\pi}}\phi$ and the potential $V\to \sqrt{2\pi} V$ to get rid of a $4\pi$ factor in the coupling constant, thereby obtaining from \eqref{eq:EFE}  \begin{subequations}
\begin{align}
r,_{uv}&=-\frac1r r,_u r,_v-\frac1{4r}\Omega^2+\frac12 r\Omega^2 V(\phi),
\label{eq:EFEdn1}\\
(\log\Omega),_{uv}&=\left(\frac{\Omega}{2r}\right)^2+\frac{r,_ur,_v}{r^2}-\phi,_u\phi,_v\label{eq:EFEdn2}\\
\phi,_{uv}&=-\frac{\phi,_ur,_v+\phi,_vr,_u}{r}-\frac14\Omega^2 V'(\phi),\label{eq:EFEdn3}
\end{align}
as a system of independent PDEs. Moreover, the two relations
\begin{align}
\left(\frac{r,_u}{\Omega^2}\right),_u&=-r\left(\frac{\phi,_u}{\Omega}\right)^2,\\
\left(\frac{r,_v}{\Omega^2}\right),_v&=-r\left(\frac{\phi,_v}{\Omega}\right)^2\label{eq:Guv}
\end{align}
\end{subequations}
are consequence of \eqref{eq:EFEdn1}--\eqref{eq:EFEdn3}. 

It will also be useful to consider the Misner--Sharp mass\footnote{often referred to as \textit{Hawking} mass too, see e.g. \cite{Dafermos:2004ws}.} $m(u,v)$ defined by the relation
\begin{equation}
\label{eq:mass}
1-\frac{2m}{r}=g(\nabla r,\nabla r),
\end{equation}
that in view of \eqref{eq:dnull} becomes
\begin{equation}
\label{eq:mass-dn}
m(u,v)=\frac r2\left(1+4\frac{r,_ur,_v}{\Omega^2}\right)
\end{equation}

\subsection{Trapped region formation in the free massless case}\label{sec:trapped}
As already mentioned, Christodoulou considers the free massless case, $V(\phi)=0$, that implies (see \eqref{eq:Bianchi}) than the scalar field satisfies the homogeneous wave equation.  The equations \eqref{eq:EFEdn1}--\eqref{eq:Guv} can be transformed in a first order (overdetermined) system 
with the positions
\begin{equation}\label{eq:cov}
\nu=r,_u,\quad\lambda=r,_v,\quad\zeta=r\phi,_u,\quad\theta=r\phi,_v
\end{equation}
in the unknown functions ($r,\lambda,\nu,m,\zeta,\theta$) as follows (see e.g. \cite[eqns. (12)--(19)]{Dafermos:2003qz})
\begin{subequations}
\begin{align}
r,_u&=\nu, &
r,_v&=\lambda,\\
\lambda,_u&=\lambda\left(\frac{2m\,\nu}{r^2-2m\,r}\right), &
\nu,_v&=\nu\left(\frac{2m\,\lambda}{r^2-2m\,r}\right),\\
m,_u&=\frac{(r-2m)\zeta^2}{2r\,\nu}, &
m,_v&=\frac{(r-2m)\theta^2}{2r\,\lambda},\\
\theta,_u&=-\frac{\zeta\lambda}{r}, &
\zeta,_v&=-\frac{\theta\nu}{r}.
\end{align}
\end{subequations}

The first paper on the subject, as already recalled, is \cite{Christodoulou:1986zr}, where a condition on the initial data is derived to have global forward--in--time solutions. This result was recently refined in \cite{Luk:2014sha,Luk2018}, where it is proven that initial data can be chosen such that their $L^1$ norm is infinite but they evolve again to a forward--in--time global solution. The construction can be extended in order to prescribe initial data at past null infinity. Moreover these refinements contain estimates useful for proving recent stability results of dispersive spherical solutions with respect to non--spherical perturbations \cite{Luk:2021ffc,Kilgore:2021loy}.

On the other side, on  \cite{Christodoulou:1991yfa}  a sufficient condition for trapped surface formation starting from collapsing initial data is formulated. The theorem is stated -- and re--proved -- in terms of the double null framework sketched above in the recent paper \cite{An:2020vdf}, that here we follow. 

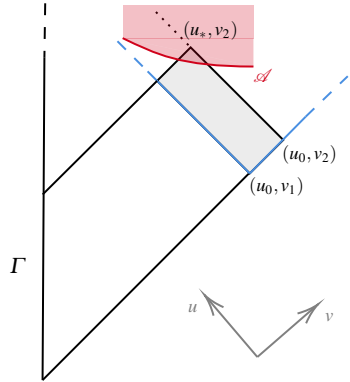
\begin{figure}
\centering

\tikzset{every picture/.style={line width=0.75pt}} 

\begin{tikzpicture}[x=0.75pt,y=0.75pt,yscale=-1,xscale=1]

\draw    (176.67,242) -- (177.6,79) ;
\draw    (176.67,242) -- (280.97,137.47) ;
\draw [color={rgb, 255:red, 74; green, 144; blue, 226 }  ,draw opacity=1 ] [dash pattern={on 4.5pt off 4.5pt}]  (309.4,109) -- (321.42,97.38) -- (330.6,88.6) ;
\draw  [color={rgb, 255:red, 0; green, 0; blue, 0 }  ,draw opacity=1 ][fill={rgb, 255:red, 155; green, 155; blue, 155 }  ,fill opacity=0.19 ] (251.35,74.05) -- (298.07,120.77) -- (281.16,137.67) -- (234.44,90.95) -- cycle ;
\draw [color={rgb, 255:red, 74; green, 144; blue, 226 }  ,draw opacity=1 ]   (229.64,86.15) -- (281.16,137.67) ;
\draw  [dash pattern={on 0.84pt off 2.51pt}]  (233.85,56.18) -- (251.35,74.05) ;
\draw [color={rgb, 255:red, 208; green, 2; blue, 27 }  ,draw opacity=1 ][fill={rgb, 255:red, 208; green, 2; blue, 27 }  ,fill opacity=0.25 ]   (217.05,69.38) .. controls (233.85,77.38) and (245.83,83.72) .. (283.03,83.72) ;
\draw    (176.73,148.29) -- (234.44,90.95) ;
\draw [color={rgb, 255:red, 128; green, 128; blue, 128 }  ,draw opacity=1 ]   (285,230.33) -- (259.63,200.52) ;
\draw [shift={(258.33,199)}, rotate = 49.6] [color={rgb, 255:red, 128; green, 128; blue, 128 }  ,draw opacity=1 ][line width=0.75]    (10.93,-3.29) .. controls (6.95,-1.4) and (3.31,-0.3) .. (0,0) .. controls (3.31,0.3) and (6.95,1.4) .. (10.93,3.29)   ;
\draw [color={rgb, 255:red, 128; green, 128; blue, 128 }  ,draw opacity=1 ]   (285,230.33) -- (314.16,204.98) ;
\draw [shift={(315.67,203.67)}, rotate = 138.99] [color={rgb, 255:red, 128; green, 128; blue, 128 }  ,draw opacity=1 ][line width=0.75]    (10.93,-3.29) .. controls (6.95,-1.4) and (3.31,-0.3) .. (0,0) .. controls (3.31,0.3) and (6.95,1.4) .. (10.93,3.29)   ;
\draw [color={rgb, 255:red, 74; green, 144; blue, 226 }  ,draw opacity=1 ]   (309.4,109) -- (281.16,137.67) ;
\draw  [draw opacity=0][fill={rgb, 255:red, 208; green, 2; blue, 27 }  ,fill opacity=0.25 ] (283.03,83.72) -- (217.05,69.38) -- (283.03,69.38) -- cycle ;
\draw  [draw opacity=0][fill={rgb, 255:red, 208; green, 2; blue, 27 }  ,fill opacity=0.25 ] (217.05,52.6) -- (283.03,52.6) -- (283.03,69.38) -- (217.05,69.38) -- cycle ;
\draw [color={rgb, 255:red, 74; green, 144; blue, 226 }  ,draw opacity=1 ] [dash pattern={on 4.5pt off 4.5pt}]  (214.44,70.95) -- (229.64,86.15) ;
\draw  [dash pattern={on 4.5pt off 4.5pt}]  (178,51.8) -- (177.6,79) ;

\draw (279.37,138.07) node [anchor=north west][inner sep=0.75pt]  [font=\scriptsize] [align=left] {$\displaystyle ( u_{0} ,v_{1})$};
\draw (159,179) node [anchor=north west][inner sep=0.75pt]  [xslant=0] [align=left] {$\displaystyle \Gamma $};
\draw (281.53,83.53) node [anchor=north west][inner sep=0.75pt]  [font=\scriptsize,color={rgb, 255:red, 208; green, 2; blue, 27 }  ,opacity=1 ] [align=left] {$\displaystyle \mathcal{A}$};
\draw (296.07,121.77) node [anchor=north west][inner sep=0.75pt]  [font=\scriptsize] [align=left] {$\displaystyle ( u_{0} ,v_{2})$};
\draw (246.07,59.37) node [anchor=north west][inner sep=0.75pt]  [font=\scriptsize] [align=left] {$\displaystyle ( u_{*} ,v_{2})$};
\draw (256.33,202) node [anchor=north east] [inner sep=0.75pt]  [font=\scriptsize,color={rgb, 255:red, 128; green, 128; blue, 128 }  ,opacity=1 ] [align=left] {$\displaystyle u$};
\draw (317.67,206.67) node [anchor=north west][inner sep=0.75pt]  [font=\scriptsize,color={rgb, 255:red, 128; green, 128; blue, 128 }  ,opacity=1 ] [align=left] {$\displaystyle v$};

\end{tikzpicture}

\caption{The collapse theorem in \cite{Christodoulou:1991yfa}. The centre of symmetry is $\Gamma$, and  initial data are prescribed in $u=u_0$ and $v=v_1$ (blue lines). A trapped region (in red), bounded by the apparent horizon $\mathcal A$, develops in the (grey) shaded region.}
\label{fig:chr1}
\end{figure}

First of all, notice that spherical symmetry allows to consider a timelike curve $\Gamma$ that will play the role of the symmetry centre. Characteristic initial data along $u=u_0$ and $v=v_1$ are prescribed (see Figure \ref{fig:chr1}), and $v_2>v_1$ is taken to define the quantities
$$
\eta_0:=\frac{m(u_0,v_2)-m(u_0,v_1)}{r(u_0,v_2)},\qquad \delta_0:=2
\frac{r(u_0,v_2)-r(u_0,v_1)}{r(u_0,v_2)}, 
$$

Now, a collapsing situation amounts to assume $r,_u<0$ and then, recalling \eqref{eq:mass-dn}, trapped surface forms when also the other partial derivative becomes negative somewhere, $r,_v<0$.

\begin{theorem}[\cite{Christodoulou:1991yfa}, Theorem 5.1]
There exists a constant $c_1\ge 1$ such that if 
$$
\delta_0\le \frac 1e\quad\text{and}\quad\eta_0>c_1 \delta_0\log\left(\frac1{\delta_0}\right)
$$ 
then
a trapped surface forms in the region $[u_0,u_*]\times[v_1,v_2]$, where $u_*$ is  such that $r(u_*,v_2)=\tfrac{3\delta_0}{1+\delta_0}$. 
\label{thm:chr1}
\end{theorem}

It is worth mentioning here that the techniques
pioneered in \cite{Christodoulou:1991yfa} to prove the above Theorem lie at the foundations of many other important  results in the gravitational collapse mathematical theory, let me cite the celebrated \cite{Christodoulou:2008nj} proof of trapped surface development from a packet of waves emitted at past infinity. There, the technique is refined to assume that the incoming radiation per unit solid angle is uniformly bounded from below. The original idea is contained in the above Theorem: the conditions on $\eta_0$ and $\delta_0$ in Theorem \ref{thm:chr1} statement \footnote{In the original formulation of \cite{Christodoulou:1991yfa}, the conditions were stated as $\delta_0<\tfrac12$ and $\eta_0>E(\delta_0)$, with  $E(x)=\frac{x}{(x+1)^2}\left(5-x-\log(2x)\right)
$.  The formulation shown here is introduced by the Author since \cite{Chr99a}, where the result is used for the proof of  naked singularity instability (see below).
}  are roughly equivalent to assuming that a sufficient amount of energy is emitted by the scalar field, at a sufficiently short amount of time.

\begin{figure}
\centering

\tikzset{every picture/.style={line width=0.75pt}} 

\begin{tikzpicture}[x=0.75pt,y=0.75pt,yscale=-1,xscale=1]

\draw    (149.67,290) -- (149.6,93.8) ;
\draw    (149.67,290) -- (253.97,185.47) ;
\draw [color={rgb, 255:red, 74; green, 144; blue, 226 }  ,draw opacity=1 ] [dash pattern={on 4.5pt off 4.5pt}]  (282.4,157) -- (294.42,145.38) -- (303.6,136.6) ;
\draw  [color={rgb, 255:red, 0; green, 0; blue, 0 }  ,draw opacity=1 ][fill={rgb, 255:red, 155; green, 155; blue, 155 }  ,fill opacity=0.19 ] (224.35,122.05) -- (271.07,168.77) -- (254.16,185.67) -- (207.44,138.95) -- cycle ;
\draw [color={rgb, 255:red, 74; green, 144; blue, 226 }  ,draw opacity=1 ]   (202.64,134.15) -- (254.16,185.67) ;
\draw  [dash pattern={on 0.84pt off 2.51pt}]  (206.85,104.18) -- (224.35,122.05) ;
\draw [color={rgb, 255:red, 208; green, 2; blue, 27 }  ,draw opacity=1 ][fill={rgb, 255:red, 208; green, 2; blue, 27 }  ,fill opacity=0.25 ]   (190.05,117.38) .. controls (206.85,125.38) and (218.83,131.72) .. (256.03,131.72) ;
\draw    (149.73,196.29) -- (207.44,138.95) ;
\draw [color={rgb, 255:red, 128; green, 128; blue, 128 }  ,draw opacity=1 ]   (258,278.33) -- (232.63,248.52) ;
\draw [shift={(231.33,247)}, rotate = 49.6] [color={rgb, 255:red, 128; green, 128; blue, 128 }  ,draw opacity=1 ][line width=0.75]    (10.93,-3.29) .. controls (6.95,-1.4) and (3.31,-0.3) .. (0,0) .. controls (3.31,0.3) and (6.95,1.4) .. (10.93,3.29)   ;
\draw [color={rgb, 255:red, 128; green, 128; blue, 128 }  ,draw opacity=1 ]   (258,278.33) -- (287.16,252.98) ;
\draw [shift={(288.67,251.67)}, rotate = 138.99] [color={rgb, 255:red, 128; green, 128; blue, 128 }  ,draw opacity=1 ][line width=0.75]    (10.93,-3.29) .. controls (6.95,-1.4) and (3.31,-0.3) .. (0,0) .. controls (3.31,0.3) and (6.95,1.4) .. (10.93,3.29)   ;
\draw [color={rgb, 255:red, 74; green, 144; blue, 226 }  ,draw opacity=1 ]   (282.4,157) -- (254.16,185.67) ;
\draw    (151.95,93.81) .. controls (153.62,92.15) and (155.29,92.16) .. (156.95,93.83) .. controls (158.61,95.5) and (160.28,95.51) .. (161.95,93.85) .. controls (163.62,92.19) and (165.28,92.19) .. (166.95,93.86) .. controls (168.61,95.53) and (170.28,95.54) .. (171.95,93.88) .. controls (173.62,92.22) and (175.29,92.23) .. (176.95,93.9) .. controls (178.61,95.57) and (180.28,95.58) .. (181.95,93.92) .. controls (183.62,92.26) and (185.29,92.27) .. (186.95,93.94) .. controls (188.61,95.61) and (190.28,95.62) .. (191.95,93.96) .. controls (193.62,92.3) and (195.29,92.31) .. (196.95,93.98) .. controls (198.61,95.65) and (200.28,95.66) .. (201.95,94) .. controls (203.62,92.34) and (205.28,92.34) .. (206.95,94.01) .. controls (208.61,95.68) and (210.28,95.69) .. (211.95,94.03) .. controls (213.62,92.37) and (215.29,92.38) .. (216.95,94.05) .. controls (218.61,95.72) and (220.28,95.73) .. (221.95,94.07) .. controls (223.62,92.41) and (225.29,92.42) .. (226.95,94.09) .. controls (228.61,95.76) and (230.28,95.77) .. (231.95,94.11) .. controls (233.62,92.45) and (235.29,92.46) .. (236.95,94.13) .. controls (238.62,95.8) and (240.28,95.8) .. (241.95,94.14) .. controls (243.62,92.48) and (245.29,92.49) .. (246.95,94.16) .. controls (248.61,95.83) and (250.28,95.84) .. (251.95,94.18) -- (256.8,94.2) -- (256.8,94.2) ;
\draw [shift={(149.6,93.8)}, rotate = 0.21] [color={rgb, 255:red, 0; green, 0; blue, 0 }  ][line width=0.75]      (0, 0) circle [x radius= 3.35, y radius= 3.35]   ;
\draw [color={rgb, 255:red, 208; green, 2; blue, 27 }  ,draw opacity=1 ][fill={rgb, 255:red, 208; green, 2; blue, 27 }  ,fill opacity=0.25 ]   (149.6,93.8) .. controls (157.84,101.2) and (172.83,106.12) .. (190.05,117.36) ;
\draw  [draw opacity=0][fill={rgb, 255:red, 208; green, 2; blue, 27 }  ,fill opacity=0.25 ] (190.05,117.38) -- (149.6,93.8) -- (190.05,93.8) -- cycle ;
\draw  [draw opacity=0][fill={rgb, 255:red, 208; green, 2; blue, 27 }  ,fill opacity=0.25 ] (256.03,131.72) -- (190.05,117.38) -- (256.03,117.38) -- cycle ;
\draw  [draw opacity=0][fill={rgb, 255:red, 208; green, 2; blue, 27 }  ,fill opacity=0.25 ] (190.05,93.8) -- (256.03,93.8) -- (256.03,117.38) -- (190.05,117.38) -- cycle ;
\draw [color={rgb, 255:red, 74; green, 144; blue, 226 }  ,draw opacity=1 ] [dash pattern={on 4.5pt off 4.5pt}]  (187.44,118.95) -- (202.64,134.15) ;
\draw [color={rgb, 255:red, 128; green, 128; blue, 128 }  ,draw opacity=1 ]   (149.6,93.8) -- (248.14,191.28) ;

\draw (252.37,186.07) node [anchor=north west][inner sep=0.75pt]  [font=\scriptsize] [align=left] {$\displaystyle ( u_{0} ,v_{1})$};
\draw (132,227) node [anchor=north west][inner sep=0.75pt]  [xslant=0] [align=left] {$\displaystyle \Gamma $};
\draw (254.53,131.53) node [anchor=north west][inner sep=0.75pt]  [font=\scriptsize,color={rgb, 255:red, 208; green, 2; blue, 27 }  ,opacity=1 ] [align=left] {$\displaystyle \mathcal{A}$};
\draw (269.07,169.77) node [anchor=north west][inner sep=0.75pt]  [font=\scriptsize] [align=left] {$\displaystyle ( u_{0} ,v_{2})$};
\draw (219.07,107.37) node [anchor=north west][inner sep=0.75pt]  [font=\scriptsize] [align=left] {$\displaystyle ( u_{*} ,v_{2})$};
\draw (229.33,250) node [anchor=north east] [inner sep=0.75pt]  [font=\scriptsize,color={rgb, 255:red, 128; green, 128; blue, 128 }  ,opacity=1 ] [align=left] {$\displaystyle u$};
\draw (290.67,254.67) node [anchor=north west][inner sep=0.75pt]  [font=\scriptsize,color={rgb, 255:red, 128; green, 128; blue, 128 }  ,opacity=1 ] [align=left] {$\displaystyle v$};
\draw (208.6,78) node [anchor=north west][inner sep=0.75pt]  [font=\footnotesize] [align=left] {$\displaystyle \mathcal{B}$};
\draw (129,89.2) node [anchor=north west][inner sep=0.75pt]  [font=\footnotesize] [align=left] {$\displaystyle \mathcal{B}_{0}$};
\draw (173.33,130.33) node [anchor=north west][inner sep=0.75pt]  [font=\footnotesize,color={rgb, 255:red, 128; green, 128; blue, 128 }  ,opacity=1 ] [align=left] {$\displaystyle \mathcal{N}$};

\end{tikzpicture}

\caption{The existence of a future singular boundary shown in \cite{Christodoulou:1991yfa}. The non-central boundary is strictly spacelike and terminates in the central boundary, where it is intersected by the apparent horizon $\mathcal A$. An incoming null hypersurface $\mathcal N$ also intersects the (singular) central boundary.} 
\label{fig:chr2}
\end{figure}
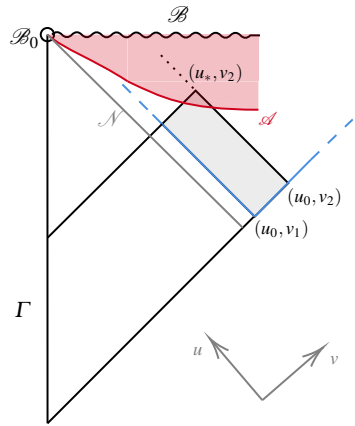

Actually, \cite[Theorem 5.1]{Christodoulou:1991yfa} contains other important results for the present context, that haven't been reported in the above statement of Theorem \ref{thm:chr1}. 
In particular, see Figure \ref{fig:chr2}, the existence of a (non-central) strictly spacelike singular future boundary is proven. This boundary terminates at the boundary of the centre of symmetry where also the apparent horizon $\mathcal A$ arrives. 
In other words, the centre of symmetry becomes trapped exactly when it becomes singular\footnote{This is a common feature shared by other models of gravitational collapse with matter, for instance collapsing stars with dust \cite{Singh:1994tb,Mena:2001dr} or anisotropic  \cite{Giambo:2002xc} matter. See also \cite{Giambo:2002tp} and references therein.}.

In a more recent paper  \cite{Dafermos:2003qz} Mihalis Dafermos will show that these solutions can be regularly extended backwards up to past null infinity. 

The paper \cite{Chr93} study solutions with bounded variation (BV) initial data. With this notation it is means that the functions $\phi,\,r,_u,\,r,_v,\,(r\phi),_u,\,(r\phi),_v$ are BV with respect to one null variable, uniformly in the other, and vice versa; plus some technical regularity conditions on the centre of symmetry $\Gamma$. The paper considers both regular solutions and those developing singularities. 

A regularity condition is stated in terms on the behaviour at the centre (see Fig. \ref{fig:chr3}): if $v_*>0$ is such that, $\forall v<v_*$, a solution exists in $\mathcal D(0,v)$, the domain of dependence of $(0,v)$, then $\exists\epsilon>0$ such that the condition 
\begin{equation}
\label{eq:ext}
\lim_{u\to v_*}\sup_{\mathcal D(u,v_*)}
\frac{2m}{r}<\epsilon
\end{equation}
implies the regular extension of the solution on $\mathcal D(0,v_1)$ for some $v_1>v_*$.

In other words, in order to extend the regular centre at some point, $2m/r$ must tend to zero as this point is approached from every direction coming from $\mathcal D(u,v_*)$. If this condition isn't satisfied then a central singularity develops, and the existence of a non-central boundary, preceded by a trapped region as in Fig. \ref{fig:chr2}, is proved.

\begin{figure}
\centering

\tikzset{every picture/.style={line width=0.75pt}} 

\begin{tikzpicture}[x=0.75pt,y=0.75pt,yscale=-1,xscale=1]

\draw    (142.78,116.42) -- (143.12,330.2) ;
\draw [shift={(142.78,116.42)}, rotate = 89.91] [color={rgb, 255:red, 0; green, 0; blue, 0 }  ][fill={rgb, 255:red, 0; green, 0; blue, 0 }  ][line width=0.75]      (0, 0) circle [x radius= 3.35, y radius= 3.35]   ;
\draw    (143.12,330.2) -- (341.97,130.83) ;
\draw  [dash pattern={on 4.5pt off 4.5pt}]  (142.78,116.42) -- (143.12,98) ;
\draw  [dash pattern={on 4.5pt off 4.5pt}]  (341.97,130.83) -- (369.5,102.61) ;
\draw [color={rgb, 255:red, 0; green, 0; blue, 0 }  ,draw opacity=1 ]   (142.78,116.42) -- (250.19,222.85) ;
\draw [color={rgb, 255:red, 208; green, 2; blue, 27 }  ,draw opacity=1 ]   (143.14,186.74) -- (179.83,152.3) ;
\draw  [draw opacity=0][fill={rgb, 255:red, 208; green, 2; blue, 27 }  ,fill opacity=0.26 ] (179.83,152.77) -- (143.14,186.74) -- (142.79,116.42) -- cycle ;
\draw [color={rgb, 255:red, 74; green, 144; blue, 226 }  ,draw opacity=1 ]   (143.12,201.38) -- (208,265.67) ;
\draw  [draw opacity=0][fill={rgb, 255:red, 74; green, 144; blue, 226 }  ,fill opacity=0.31 ] (207.58,265.67) -- (143.12,330.2) -- (143.12,201.38) -- cycle ;
\draw [color={rgb, 255:red, 74; green, 144; blue, 226 }  ,draw opacity=1 ]   (111.92,257.81) .. controls (120.91,246.39) and (128.73,236.15) .. (155.53,250.18) ;
\draw [shift={(157.19,251.06)}, rotate = 208.59] [color={rgb, 255:red, 74; green, 144; blue, 226 }  ,draw opacity=1 ][line width=0.75]    (4.37,-1.32) .. controls (2.78,-0.56) and (1.32,-0.12) .. (0,0) .. controls (1.32,0.12) and (2.78,0.56) .. (4.37,1.32)   ;
\draw [color={rgb, 255:red, 208; green, 2; blue, 27 }  ,draw opacity=1 ]   (101.52,148.01) .. controls (105.13,156.46) and (118.25,167.9) .. (152.55,148.89) ;
\draw [shift={(154.13,148.01)}, rotate = 150.3] [color={rgb, 255:red, 208; green, 2; blue, 27 }  ,draw opacity=1 ][line width=0.75]    (4.37,-1.32) .. controls (2.78,-0.56) and (1.32,-0.12) .. (0,0) .. controls (1.32,0.12) and (2.78,0.56) .. (4.37,1.32)   ;
\draw  [draw opacity=0][fill={rgb, 255:red, 155; green, 155; blue, 155 }  ,fill opacity=0.2 ][dash pattern={on 0.84pt off 2.51pt}] (248.84,222.78) -- (142.78,330.2) -- (142.78,116.42) -- cycle ;
\draw  [dash pattern={on 4.5pt off 4.5pt}]  (142.67,92.11) -- (261.61,211.19) ;
\draw [color={rgb, 255:red, 128; green, 128; blue, 128 }  ,draw opacity=1 ]   (310.43,311.15) -- (281.09,277.28) ;
\draw [shift={(279.78,275.76)}, rotate = 49.11] [color={rgb, 255:red, 128; green, 128; blue, 128 }  ,draw opacity=1 ][line width=0.75]    (10.93,-3.29) .. controls (6.95,-1.4) and (3.31,-0.3) .. (0,0) .. controls (3.31,0.3) and (6.95,1.4) .. (10.93,3.29)   ;
\draw [color={rgb, 255:red, 128; green, 128; blue, 128 }  ,draw opacity=1 ]   (310.43,311.15) -- (344.15,282.33) ;
\draw [shift={(345.67,281.03)}, rotate = 139.49] [color={rgb, 255:red, 128; green, 128; blue, 128 }  ,draw opacity=1 ][line width=0.75]    (10.93,-3.29) .. controls (6.95,-1.4) and (3.31,-0.3) .. (0,0) .. controls (3.31,0.3) and (6.95,1.4) .. (10.93,3.29)   ;
\draw  [draw opacity=0][fill={rgb, 255:red, 184; green, 233; blue, 134 }  ,fill opacity=0.3 ] (154.81,104.39) -- (261.73,211.31) -- (250.19,222.85) -- (143.27,115.93) -- cycle ;
\draw  [draw opacity=0][fill={rgb, 255:red, 184; green, 233; blue, 134 }  ,fill opacity=0.3 ] (144.07,115.13) -- (142.67,92.11) -- (154.88,104.32) -- cycle ;

\draw (250.84,225.78) node [anchor=north west][inner sep=0.75pt]  [font=\scriptsize] [align=left] {$\displaystyle ( 0,v_{*})$};
\draw (209.58,268.67) node [anchor=north west][inner sep=0.75pt]  [font=\scriptsize] [align=left] {$\displaystyle ( 0,v)$};
\draw (89.36,263.27) node [anchor=north west][inner sep=0.75pt]  [font=\footnotesize,color={rgb, 255:red, 74; green, 122; blue, 226 }  ,opacity=1 ] [align=left] {$\displaystyle \mathcal{D}( 0,v)$};
\draw (88.24,133.21) node [anchor=north west][inner sep=0.75pt]  [font=\footnotesize,color={rgb, 255:red, 208; green, 2; blue, 27 }  ,opacity=1 ] [align=left] {$\displaystyle \mathcal{D}( u,v_{*})$};
\draw (263.61,214.19) node [anchor=north west][inner sep=0.75pt]  [font=\scriptsize] [align=left] {$\displaystyle ( 0,v_{1})$};
\draw (125.06,196.19) node [anchor=north west][inner sep=0.75pt]  [font=\footnotesize] [align=left] {$\displaystyle \Gamma $};
\draw (276.66,279.93) node [anchor=north east] [inner sep=0.75pt]  [font=\scriptsize,color={rgb, 255:red, 128; green, 128; blue, 128 }  ,opacity=1 ] [align=left] {$\displaystyle u$};
\draw (348.79,285.2) node [anchor=north west][inner sep=0.75pt]  [font=\scriptsize,color={rgb, 255:red, 128; green, 128; blue, 128 }  ,opacity=1 ] [align=left] {$\displaystyle v$};

\end{tikzpicture}

\caption{The extension theorem from \cite{Chr93}.
Condition \eqref{eq:ext} allows to extend the solution to a  region (in green) across the future null cone of $(0,v_*)$. 
}
\label{fig:chr3}
\end{figure}

\subsection{Non-genericity of naked singularity}\label{sec:instability}

As said in the introduction, the solution exhibiting a naked singularity is built in the subsequent paper \cite{Christodoulou:1994hg}. It is better to recall here that the naked singularity here corresponds to a situation where horizon does not exist at all\footnote{In this sense this situation is very different from the naked singularity developing in the collapse of some matter models, see previous footnote. In those contexts there may be situations where outgoing null geodesics are ``emitted'' from the central singularity and detected by a faraway observer, even when a horizon forms.}. 
We will discuss this example in next Section, since it is in some way connected to the critical solution numerically found by Choptuik \cite{Choptuik:1992jv}. 
Here we only stress that this example exhibits a pointwise singularity on the centre of symmetry. This fact is relevant because, in \cite{Chr99a}, it is shown that if one has bounded variation initial data on $u=0$ developing such a solution, or more in general a solution that is neither future regular nor develops an event horizon, then a suitable perturbation of this initial data restores the apparent horizon and, hence, trapped surface formation. 

More specifically, assigned  BV   initial data  such that a naked pointwise singularity develops at the centre, then two BV functions $f_1(v),\,f_2(v)$ exist, such that \textit{any} (non-trivial) perturbation of the initial data with a linear combination of $f_1(v)$ and $f_2(v)$ results in a trapped surface formation.
Let us discuss in brief this important result.

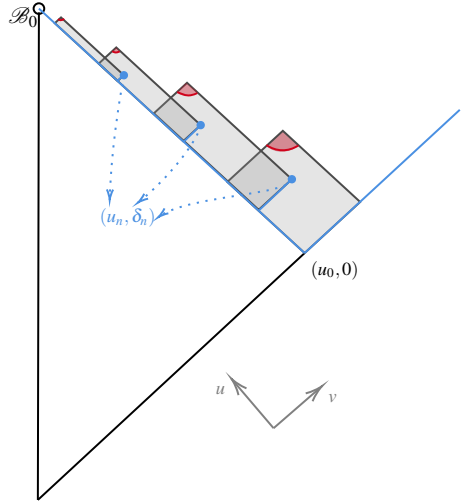
\begin{figure}
\centering

\tikzset{every picture/.style={line width=0.75pt}} 

\begin{tikzpicture}[x=0.75pt,y=0.75pt,yscale=-0.8,xscale=0.8]

\draw  [draw opacity=0][fill={rgb, 255:red, 208; green, 2; blue, 27 }  ,fill opacity=0.48 ] (264.66,96.9) .. controls (258.92,102.11) and (249.9,102.28) .. (244.45,97.26) -- (254.84,87.7) -- cycle ;
\draw    (101.06,12.95) -- (100.4,320.59) ;
\draw [shift={(101.07,10.6)}, rotate = 90.12] [color={rgb, 255:red, 0; green, 0; blue, 0 }  ][line width=0.75]      (0, 0) circle [x radius= 3.35, y radius= 3.35]   ;
\draw    (100.4,320.59) -- (268.58,164.83) ;
\draw  [color={rgb, 255:red, 74; green, 74; blue, 74 }  ,draw opacity=1 ][fill={rgb, 255:red, 155; green, 155; blue, 155 }  ,fill opacity=0.26 ] (254.84,87.7) -- (303.59,132.59) -- (268.58,164.83) -- (219.83,119.94) -- cycle ;
\draw  [color={rgb, 255:red, 74; green, 74; blue, 74 }  ,draw opacity=1 ][fill={rgb, 255:red, 155; green, 155; blue, 155 }  ,fill opacity=0.26 ] (194.45,57.4) -- (260.53,118.26) -- (239.61,137.52) -- (173.53,76.66) -- cycle ;
\draw [color={rgb, 255:red, 74; green, 144; blue, 226 }  ,draw opacity=1 ]   (260.54,118.25) -- (239.9,137.78) ;
\draw [shift={(260.54,118.25)}, rotate = 136.57] [color={rgb, 255:red, 74; green, 144; blue, 226 }  ,draw opacity=1 ][fill={rgb, 255:red, 74; green, 144; blue, 226 }  ,fill opacity=1 ][line width=0.75]      (0, 0) circle [x radius= 2.01, y radius= 2.01]   ;
\draw  [color={rgb, 255:red, 74; green, 74; blue, 74 }  ,draw opacity=1 ][fill={rgb, 255:red, 155; green, 155; blue, 155 }  ,fill opacity=0.26 ] (149.83,35.1) -- (203.05,84.11) -- (192,94.29) -- (138.77,45.28) -- cycle ;
\draw  [color={rgb, 255:red, 74; green, 74; blue, 74 }  ,draw opacity=1 ][fill={rgb, 255:red, 155; green, 155; blue, 155 }  ,fill opacity=0.26 ] (115.01,16.17) -- (154.62,52.64) -- (150.68,56.27) -- (111.07,19.8) -- cycle ;
\draw [color={rgb, 255:red, 74; green, 144; blue, 226 }  ,draw opacity=1 ]   (101.07,10.6) -- (149.93,55.58) -- (268.58,164.83) ;
\draw [color={rgb, 255:red, 74; green, 144; blue, 226 }  ,draw opacity=1 ]   (268.58,164.83) -- (366.4,74.44) ;
\draw [color={rgb, 255:red, 74; green, 144; blue, 226 }  ,draw opacity=1 ]   (203.05,84.11) -- (192.33,94.6) ;
\draw [shift={(203.05,84.11)}, rotate = 135.63] [color={rgb, 255:red, 74; green, 144; blue, 226 }  ,draw opacity=1 ][fill={rgb, 255:red, 74; green, 144; blue, 226 }  ,fill opacity=1 ][line width=0.75]      (0, 0) circle [x radius= 2.01, y radius= 2.01]   ;
\draw [color={rgb, 255:red, 74; green, 144; blue, 226 }  ,draw opacity=1 ]   (154.62,52.64) -- (150.16,56.39) ;
\draw [shift={(154.62,52.64)}, rotate = 139.96] [color={rgb, 255:red, 74; green, 144; blue, 226 }  ,draw opacity=1 ][fill={rgb, 255:red, 74; green, 144; blue, 226 }  ,fill opacity=1 ][line width=0.75]      (0, 0) circle [x radius= 2.01, y radius= 2.01]   ;
\draw [color={rgb, 255:red, 208; green, 2; blue, 27 }  ,draw opacity=1 ]   (244.46,96.66) .. controls (250.56,100.87) and (258.14,101.2) .. (264.67,96.9) ;
\draw [color={rgb, 255:red, 208; green, 2; blue, 27 }  ,draw opacity=1 ]   (188.24,63.24) .. controls (193.67,65.59) and (195.61,66.32) .. (200.97,63.85) ;
\draw  [draw opacity=0][fill={rgb, 255:red, 208; green, 2; blue, 27 }  ,fill opacity=0.35 ] (200.84,62.77) .. controls (200.65,62.99) and (200.44,63.2) .. (200.22,63.41) .. controls (196.94,66.43) and (191.6,66.46) .. (188,63.58) -- (193.72,57.43) -- cycle ;
\draw [color={rgb, 255:red, 208; green, 2; blue, 27 }  ,draw opacity=1 ]   (147.26,37.46) .. controls (149.2,38.73) and (150.4,38.33) .. (152,37.53) ;
\draw  [draw opacity=0][fill={rgb, 255:red, 208; green, 2; blue, 27 }  ,fill opacity=0.35 ] (147.69,37.3) .. controls (148.08,37.59) and (148.53,37.83) .. (149.02,38.01) .. controls (150.25,38.46) and (151.47,38.42) .. (152.36,37.97) -- (149.83,35.1) -- cycle ;
\draw [color={rgb, 255:red, 208; green, 2; blue, 27 }  ,draw opacity=1 ]   (113.12,17.28) .. controls (114.67,18.33) and (115.47,17.93) .. (117.07,17.13) ;
\draw [color={rgb, 255:red, 74; green, 144; blue, 226 }  ,draw opacity=1 ] [dash pattern={on 0.84pt off 2.51pt}]  (154.62,52.64) .. controls (152.84,62.79) and (146.87,95.31) .. (145.74,129.73) ;
\draw [shift={(145.69,131.3)}, rotate = 271.64] [color={rgb, 255:red, 74; green, 144; blue, 226 }  ,draw opacity=1 ][line width=0.75]    (6.56,-1.97) .. controls (4.17,-0.84) and (1.99,-0.18) .. (0,0) .. controls (1.99,0.18) and (4.17,0.84) .. (6.56,1.97)   ;
\draw [color={rgb, 255:red, 74; green, 144; blue, 226 }  ,draw opacity=1 ] [dash pattern={on 0.84pt off 2.51pt}]  (203.05,84.11) .. controls (201.29,94.16) and (182.48,117.78) .. (164.85,131.93) ;
\draw [shift={(163.5,133)}, rotate = 322.13] [color={rgb, 255:red, 74; green, 144; blue, 226 }  ,draw opacity=1 ][line width=0.75]    (6.56,-1.97) .. controls (4.17,-0.84) and (1.99,-0.18) .. (0,0) .. controls (1.99,0.18) and (4.17,0.84) .. (6.56,1.97)   ;
\draw [color={rgb, 255:red, 74; green, 144; blue, 226 }  ,draw opacity=1 ] [dash pattern={on 0.84pt off 2.51pt}]  (260.54,118.25) .. controls (258.77,128.3) and (197.46,128.24) .. (177.72,141.2) ;
\draw [shift={(176.28,142.23)}, rotate = 322.13] [color={rgb, 255:red, 74; green, 144; blue, 226 }  ,draw opacity=1 ][line width=0.75]    (6.56,-1.97) .. controls (4.17,-0.84) and (1.99,-0.18) .. (0,0) .. controls (1.99,0.18) and (4.17,0.84) .. (6.56,1.97)   ;
\draw [color={rgb, 255:red, 128; green, 128; blue, 128 }  ,draw opacity=1 ]   (249,275.33) -- (223.63,245.52) ;
\draw [shift={(222.33,244)}, rotate = 49.6] [color={rgb, 255:red, 128; green, 128; blue, 128 }  ,draw opacity=1 ][line width=0.75]    (10.93,-3.29) .. controls (6.95,-1.4) and (3.31,-0.3) .. (0,0) .. controls (3.31,0.3) and (6.95,1.4) .. (10.93,3.29)   ;
\draw [color={rgb, 255:red, 128; green, 128; blue, 128 }  ,draw opacity=1 ]   (249,275.33) -- (278.16,249.98) ;
\draw [shift={(279.67,248.67)}, rotate = 138.99] [color={rgb, 255:red, 128; green, 128; blue, 128 }  ,draw opacity=1 ][line width=0.75]    (10.93,-3.29) .. controls (6.95,-1.4) and (3.31,-0.3) .. (0,0) .. controls (3.31,0.3) and (6.95,1.4) .. (10.93,3.29)   ;

\draw (270.58,167.83) node [anchor=north west][inner sep=0.75pt]  [font=\scriptsize] [align=left] {$\displaystyle ( u_{0} ,0)$};
\draw (80,8.2) node [anchor=north west][inner sep=0.75pt]  [font=\footnotesize] [align=left] {$\displaystyle \mathcal{B}_{0}$};
\draw (137.58,134.83) node [anchor=north west][inner sep=0.75pt]  [font=\scriptsize,color={rgb, 255:red, 74; green, 144; blue, 226 }  ,opacity=1 ] [align=left] {$\displaystyle ( u_{n} ,\delta _{n})$};
\draw (220.33,247) node [anchor=north east] [inner sep=0.75pt]  [font=\scriptsize,color={rgb, 255:red, 128; green, 128; blue, 128 }  ,opacity=1 ] [align=left] {$\displaystyle u$};
\draw (281.67,251.67) node [anchor=north west][inner sep=0.75pt]  [font=\scriptsize,color={rgb, 255:red, 128; green, 128; blue, 128 }  ,opacity=1 ] [align=left] {$\displaystyle v$};

\end{tikzpicture}

\caption{The instability theorem in \cite{Chr99a}, in the double null setting argument proved in  \cite{Liu:2017itp}. Here $\mathcal B_0$ has coordinates $(0,0)$. The sequence $(u_n,\delta_n)$ tends to $(0^-,0^+)$ and each point is taken to satisfy hypotheses of Theorem \ref{thm:chr1} (on the blue lines) and find out a trapped region (in grey) such that a trapped surface (in red) develops in the upper corner at each step.}
\label{fig:chr4}
\end{figure}

First of all, it must be said that the proof in \cite{Chr99a} is carried out  again in a Bondi coordinate setting similar to \eqref{eq:Bondi1}. 
However, in recent times a new proof appeared \cite{Liu:2017itp} where the argument is adapted in double null coordinates. 
The rough idea of the argument is sketched in Fig. \ref{fig:chr4}.

Coordinates are shifted in such a way that $(0,0)$ is an isolated singularity in the $(u,v)$--plane as a development of some particular initial data $\alpha_0(v)=r\phi,_v(u_0,v)$. 

The idea follows these steps:
\begin{enumerate}
\item  finding proper estimates of the quantities involved (not satisfied with the initial data $\alpha_0(v)$), in order to determine a sequence $(u_n,\delta_n)$ approaching $(0,0)$, in such a way that one can apply  Theorem \ref{thm:chr1} with $v_1=0$ and $v_2=\delta_n$, and determine trapped surfaces existence. At every step of the sequence one is able to determine a piece of the apparent horizon. Since the sequence approaches the singular point, so does the apparent horizon, and the singularity is no more naked.
\item finding conditions on $r \phi,_v(0,v)$ such that estimates cited on step 1 hold. This is done by considering two separate cases, see \cite[Theorem 5 and Theorem 6]{Liu:2017itp}, corresponding to \cite[Theorem 2.1 and Theorem 3.1]{Chr99a}.
\item determining -- see \cite[Section 4]{Chr99a} and, in its new version, \cite[Proof of Theorem 3]{Liu:2017itp} -- two suitable functions $f_1(v)$ and $f_2(v)$ and proving that the initial data $\alpha_0(v)+\lambda_1 f_1(v)+\lambda_2 f_2(v)$ ($\forall\lambda_1,\lambda_2\in\mathbb R$) evolves in such a way that  $r \phi,_v(0,v)$  necessarily satisfies the conditions stated on step 2. 
\end{enumerate}

The instability result in \cite{Chr99a} marks an important point in the history of Penrose cosmic censorship conjecture, and many  still regard that paper as ``the proof'' of this conjecture, or at least of a re--designed version of it, allowing for naked singularities at most for an \textit{exceptional} set of zero measure in the set of initial data for the evolution problem. The result shows that the codimension of the space of initial data is at least 2, being at least two ``directions'' where an initial data developing a point singularity can be perturbed, to restore trapped surface formation. 

This and other results in this context seems to highlight a non-genericity feature of pointlike singularities developing without an horizon, but a central question may be the choice of the suitable -- under a physical point of view -- regularity requested for the initial data, because demanding a mild regularity property -- as bounded variation solution may be, as commented in \cite[Section 3.4]{Martin-Garcia:2003xgm}-- can result in a too large space of admissible initial data. 

In this sense, if we embrace the evolutionary PDE setting that is the object of the present section, and therefore identify the double null  as the natural coordinate setting to work with, taking initial data on a null hypersurface, it must be said, regarding the instability theorem proved in \cite{Chr99a} or in its new version of \cite{Liu:2017itp}, that at least the function $f_2(v)$ is absolutely continuous in $[0,+\infty[$ (whereas $f_1$ has a jump discontinuity). This means that non-genericity (with codimension at least 1) is granted when one restricts to absolutely continuous functions.

Of course, the scalar field model treated so far is not the most general one (recall that potential is been set to zero so far), and then in the following subsection we wish to explore some more recent result that aim to extend  the knowledge on the subject through analytical studies.

\subsection{Cosmological constant and potential}
Theorem \ref{thm:chr1} has been proved in other contexts. For example, in \cite{Costa:2020sha} the proof is made where the addition of a positive cosmological constant is considered, thereby obtaining the modified system with respect to \eqref{eq:EFE}
\begin{equation}
\label{eq:EFE1}
R_{\mu\nu}-\frac12 g_{\mu\nu}R+\Lambda g_{\mu\nu}=8\pi T_{\mu\nu}.
\end{equation}
This amounts to work with system given by \eqref{eq:EFE}, where the potential $V(\phi)$ is assumed to be constant and positive.
Therefore, with the same normalization of the scalar field considered above (see right after \eqref{eq:dnull}) we arrive to the system \eqref{eq:EFEdn1}--\eqref{eq:EFEdn3} with $V(\phi)=\Lambda/2$. The conditions are similar to those appearing in \ref{thm:chr1}, paying attention to that the introduction of $\Lambda$ affects the definition of mass \eqref{eq:mass-dn} that in this context must be normalized as $\varpi=m-\tfrac{\Lambda}{6} r^3$. 

The case with non-zero potential has in general been less covered in the PDE approach. Introducing a potential in principle adds a source term to Klein--Gordon equation and may significantly complicate the evolution in that, just to mention, scale invariance does not hold anymore. However some results have been proved all the same. Among these it must be certainly mentioned one of Dafermos paper on the subject, i.e. \cite{Dafermos:2004ws}. At the time, paper \cite{Hertog:2003zs} appeared that -- heuristically trying to approximate the model with the anti--deSitter homogeneous spacetime,  corresponding to the asymptotic state of a negative minima for the potential $V(\phi)$, was conjecturing the appearance of naked singularities generically arising but not from the centre of symmetry of the system.  The paper \cite{Dafermos:2004ws} shows that this situation cannot happen for a collapsing scalar field with bounded from below potential, regardless of the sign of $V(\phi)$ at the minima. 

\begin{figure}
\centering

\tikzset{every picture/.style={line width=0.75pt}} 

\begin{tikzpicture}[x=0.75pt,y=0.75pt,yscale=-1,xscale=1]

\draw    (100.45,112.01) -- (100.5,282) ;
\draw    (100.5,282) .. controls (140.5,252) and (359.55,300.85) .. (399.55,270.85) ;
\draw  [dash pattern={on 4.5pt off 4.5pt}]  (239.17,110.58) -- (397.89,269.19) ;
\draw [shift={(399.55,270.85)}, rotate = 44.98] [color={rgb, 255:red, 0; green, 0; blue, 0 }  ][line width=0.75]      (0, 0) circle [x radius= 3.35, y radius= 3.35]   ;
\draw  [draw opacity=0][fill={rgb, 255:red, 155; green, 155; blue, 155 }  ,fill opacity=0.35 ] (177.39,107.11) -- (226.89,156.61) -- (198.61,184.89) -- (149.11,135.39) -- cycle ;
\draw    (175.73,108.77) -- (149.11,135.39) ;
\draw [shift={(177.39,107.11)}, rotate = 135] [color={rgb, 255:red, 0; green, 0; blue, 0 }  ][line width=0.75]      (0, 0) circle [x radius= 3.35, y radius= 3.35]   ;
\draw    (177.39,107.11) -- (226.89,156.61) ;
\draw [color={rgb, 255:red, 74; green, 144; blue, 226 }  ,draw opacity=1 ][line width=1.5]    (149.11,135.39) -- (198.61,184.89) ;
\draw [color={rgb, 255:red, 74; green, 144; blue, 226 }  ,draw opacity=1 ][line width=1.5]    (226.89,156.61) -- (198.61,184.89) ;kerr 
\draw [color={rgb, 255:red, 74; green, 144; blue, 226 }  ,draw opacity=1 ] [dash pattern={on 0.84pt off 2.51pt}]  (165.5,158) .. controls (161.56,165.88) and (158.59,185.4) .. (187.17,202.23) ;
\draw [shift={(188.5,203)}, rotate = 209.54] [color={rgb, 255:red, 74; green, 144; blue, 226 }  ,draw opacity=1 ][line width=0.75]    (6.56,-1.97) .. controls (4.17,-0.84) and (1.99,-0.18) .. (0,0) .. controls (1.99,0.18) and (4.17,0.84) .. (6.56,1.97)   ;
\draw [color={rgb, 255:red, 74; green, 144; blue, 226 }  ,draw opacity=1 ] [dash pattern={on 0.84pt off 2.51pt}]  (218.5,170) .. controls (225.33,174.88) and (224.55,190.21) .. (205.96,203.95) ;
\draw [shift={(204.5,205)}, rotate = 325.01] [color={rgb, 255:red, 74; green, 144; blue, 226 }  ,draw opacity=1 ][line width=0.75]    (6.56,-1.97) .. controls (4.17,-0.84) and (1.99,-0.18) .. (0,0) .. controls (1.99,0.18) and (4.17,0.84) .. (6.56,1.97)   ;
\draw [color={rgb, 255:red, 208; green, 2; blue, 27 }  ,draw opacity=1 ][fill={rgb, 255:red, 208; green, 2; blue, 27 }  ,fill opacity=0.4 ]   (145.69,116.92) .. controls (151.94,120) and (191.17,124.67) .. (207.69,121.92) ;
\draw  [draw opacity=0][fill={rgb, 255:red, 208; green, 2; blue, 27 }  ,fill opacity=0.4 ] (207.69,121.92) -- (145.69,116.92) -- (207.69,116.92) -- cycle ;
\draw  [draw opacity=0][fill={rgb, 255:red, 208; green, 2; blue, 27 }  ,fill opacity=0.4 ] (145.69,109.07) .. controls (145.69,107.99) and (146.57,107.11) .. (147.65,107.11) -- (205.72,107.11) .. controls (206.81,107.11) and (207.69,107.99) .. (207.69,109.07) -- (207.69,116.92) .. controls (207.69,116.92) and (207.69,116.92) .. (207.69,116.92) -- (145.69,116.92) .. controls (145.69,116.92) and (145.69,116.92) .. (145.69,116.92) -- cycle ;

\draw (79,173) node [anchor=north west][inner sep=0.75pt]  [font=\small] [align=left] {$\displaystyle \Gamma $};
\draw (316,160) node [anchor=north west][inner sep=0.75pt]  [font=\small] [align=left] {$\mathcal I^{+}$};
\draw (190.5,206) node [anchor=north west][inner sep=0.75pt]  [color={rgb, 255:red, 74; green, 144; blue, 226 }  ,opacity=1 ] [align=left] {$\displaystyle X$};
\draw (167,88) node [anchor=north west][inner sep=0.75pt]  [font=\small] [align=left] {$\displaystyle p$};
\draw (401.55,273.85) node [anchor=north west][inner sep=0.75pt]   [align=left] {$\displaystyle i^{0}$};
\draw (221,277) node [anchor=north west][inner sep=0.75pt]   [align=left] {$\displaystyle S$};
\draw (205.69,118.92) node [anchor=south east] [inner sep=0.75pt]  [color={rgb, 255:red, 208; green, 2; blue, 27 }  ,opacity=1 ] [align=left] {$\mathcal T$};
\draw (143.69,119.92) node [anchor=north east] [inner sep=0.75pt]  [color={rgb, 255:red, 208; green, 2; blue, 27 }  ,opacity=1 ] [align=left] {$\mathcal A$};

\end{tikzpicture}
\caption{The result in \cite{Dafermos:2004ws}. The grey region is $J^-(p)\cap D^+(X)$. If $V(\phi)$ is bounded from below -- and then possibly attaining negative values -- then a non-central singularity is necessarily hidden inside a trapped surface $\mathcal{T}$, bounded by an apparent horizon $\mathcal A$. }
\label{fig:daf1}
\end{figure}
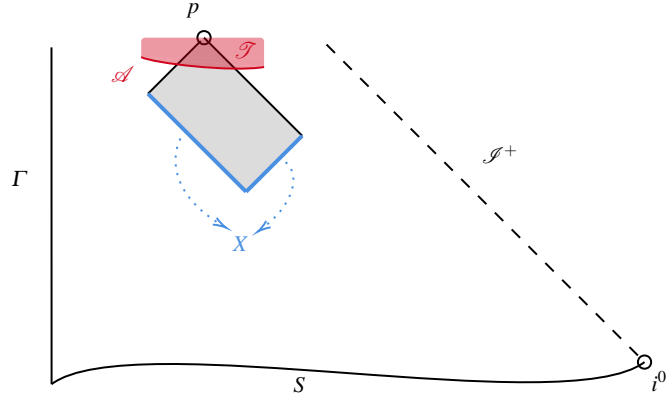

In particular, the situation described in \cite{Dafermos:2004ws} is the following. We consider the future evolution $\mathcal Q^+$ 
from a spacelike Cauchy hypersurface $S$ under the collapsing assumption $r,_u<0$ in $\mathcal Q^+$. 
The notion of \textit{first singularity} is introduced as follows:
\begin{definition}
A boundary point $p\in\partial\mathcal Q^+$ is a \textit{first singularity} if
\begin{enumerate}
\item $J^+(p)\cap\mathcal Q^+$ is \textit{eventually compactly generated} (i.e. $\exists X\subset\mathcal Q^+$ compact such that $J^-(p)\subset D^+(X)\cup J^-(X)$);
\item if $Y\subsetneq J^-(p)\cap{\mathcal Q^+}$ is eventually compactly generated $\Rightarrow\exists q\in\mathcal Q^+\,:\,Y=J^-(q)$. 
\end{enumerate}
\end{definition}
The result proved in \cite{Dafermos:2004ws}  is that (see Fig. \ref{fig:daf1}) \textit{non central} first singularities are such that
$$
J^{-}(p)\cap \mathcal Q^+\cap  D^+(X)\cap \mathcal T\ne \emptyset,
$$
where $\mathcal T$ is the trapped surface defined by $r,_v<0$. 

This means that the horizon must form prior to the appearance of a non central singularity. The argument lies in careful estimates on the quantities involved in the ``diamond'' $J^-(p)\cap D^+(X)$, and that is why non centrality of $p$ is crucial for this proof.  

The paper \cite{Dafermos:2004wr} of the same period states general sufficient conditions on energy momentum tensor for the future null infinity  $I^+$ to be future complete when $\mathcal Q^+\setminus J (I^+)\ne\emptyset$ (i.e., when a trapped region arises) and in the scalar field collapse the result can be applied when $V(\phi)\ge 0$. Indeed, one of the sufficient conditions stated in \cite{Dafermos:2004wr} is the dominant energy condition that in this case requires $V(\phi)\ge 0$ to hold. 
In the evolutionary PDE approach, this corresponds to a reformulation of weak cosmic censorship given in \cite{Chr99}, see also \cite[Section 10.5]{Landsman:2022hrn}. It is also worth mentioning that the results proved in \cite{Dafermos:2004ws,Dafermos:2004wr}  have been extended to higher dimensional spacetimes in \cite{Langfelder:2004sk}. 

\subsection{Interaction with an electromagnetic field}\label{sec:Maxw}
Recalling the dualism between the (spherically symmetric) Reissner--Nordstr\"{o}m spacetime and the (rotating) Kerr spacetime, in an effort to step out of the spherically symmetric case, the first attempt is to consider a spherical situation when an electromagnetic field is added to the system. This leads to modify the Lagrangian \eqref{eq:Lagr} as happens in the Einstein--Maxwell--scalar model, where the energy momentum tensor $T=T^{SF}+T^{EM}$ is made by the sum of two components $T^{SF}$ given by \eqref{eq:T} and an electromagnetic field 
\begin{equation}
\label{eq:Tem}
T^{EM}_{\mu\nu}=
F_{\mu\lambda}F_{\nu\rho}g^{\lambda\rho}-\frac14 g_{\mu\nu}F_{\lambda\rho}F_{\sigma\tau}g^{\lambda\sigma}g^{\rho\tau}.
\end{equation}
In this situations the scalar field and the electromagnetic field are uncoupled, so that each component of $T$ separately satisfies Bianchi equations, i.e. the divergence of each $T^{EM}$ and $T^{SF}$ separately vanishes. Consequently, the Klein--Gordon equation still holds, together with Maxwell equations
\begin{equation}
\label{eq:Maxwell}
F^{\mu\nu}_{\quad;\nu}=0,\qquad F_{[\mu\nu,\rho]}=0, 
\end{equation}
that in view of the sphericity of the metric, gives a quite simplified form for $T^{EM}$:
\begin{equation}
\label{eq:Tem-2}
(T^{EM})^\mu_\nu=\frac{q^2}{2r^4}\cdot\text{diag}\{-1,-1,1,1\},
\end{equation}
where the constant $q$ is the \textit{charge} of the EM field. The mass $m$ given in \eqref{eq:mass} is correspondingly renormalized as $\varpi=m+\frac{q^2}{2r^2}$. 

The addition of this Maxwell field in the model, as studied by Dafermos in a series of his earlier papers \cite{Dafermos:2003vim,Dafermos:2003wr,Daf2005}, 
results in a quite different evolution of the collapse. Indeed, the future maximal development of the Cauchy problems ends at least partially in a Cauchy horizon that is not necessarily a singularity in the classical sense\footnote{Interestingly enough, adding a charge to other matter models also results in weak singularity formation such as the shell crossing developing from the collapse of charged anisotropic models studied in \cite{Cipolletta:2012pf}}. Near this region the function $r$ is positive and bounded away from zero \cite[Theorem 1]{Dafermos:2003vim}, but the renormalized mass $\varpi$ blows up the at the Cauchy horizon \cite[Theorem 2]{Dafermos:2003vim}. These facts imply \cite[Theorems 1.1]{Dafermos:2003wr} that the metric can be (not uniquely) $C^0$ extended across the Cauchy horizon. In the same paper  \cite[Theorems 1.2]{Dafermos:2003wr} a sufficient condition, based on the growth estimate of $\phi,_v$ along the event horizon, is shown to imply that $C^1$ extension is not allowed. Actually Jonathan Luk and Sung--Jin Oh observe in papers \cite{Luk2019-I,Luk2019-II} that this pointwise estimate condition is difficult to be verified; in the same works, the $C^2$ inextendibility across the Cauchy horizon is proved, supporting a $C^2$--version of strong cosmic censorship.

\begin{figure}
\centering

\tikzset{every picture/.style={line width=0.75pt}} 

\begin{tikzpicture}[x=0.75pt,y=0.75pt,yscale=-1,xscale=1]

\draw  [draw opacity=0][fill={rgb, 255:red, 155; green, 155; blue, 155 }  ,fill opacity=0.82 ] (184.94,46.78) -- (198.84,89.34) -- (170.61,61.11) -- cycle ;
\draw  [draw opacity=0][fill={rgb, 255:red, 155; green, 155; blue, 155 }  ,fill opacity=0.52 ] (198.89,89.39) -- (149.39,138.89) -- (121.11,110.61) -- (170.61,61.11) -- cycle ;
\draw  [dash pattern={on 0.84pt off 2.51pt}]  (172.27,62.77) -- (198.89,89.39) ;
\draw [shift={(170.61,61.11)}, rotate = 45] [color={rgb, 255:red, 0; green, 0; blue, 0 }  ][line width=0.75]      (0, 0) circle [x radius= 3.35, y radius= 3.35]   ;
\draw  [dash pattern={on 4.5pt off 4.5pt}]  (200.55,91.06) -- (332.25,223.5) ;
\draw [shift={(198.89,89.39)}, rotate = 45.16] [color={rgb, 255:red, 0; green, 0; blue, 0 }  ][line width=0.75]      (0, 0) circle [x radius= 3.35, y radius= 3.35]   ;
\draw [color={rgb, 255:red, 128; green, 128; blue, 128 }  ,draw opacity=1 ]   (121.11,110.61) -- (149.39,138.89) ;
\draw [color={rgb, 255:red, 0; green, 0; blue, 0 }  ,draw opacity=1 ]   (149.39,138.89) -- (198.89,89.39) ;
\draw [shift={(149.39,138.89)}, rotate = 315] [color={rgb, 255:red, 0; green, 0; blue, 0 }  ,draw opacity=1 ][fill={rgb, 255:red, 0; green, 0; blue, 0 }  ,fill opacity=1 ][line width=0.75]      (0, 0) circle [x radius= 3.35, y radius= 3.35]   ;
\draw  [dash pattern={on 0.84pt off 2.51pt}]  (121.11,110.61) -- (185.25,47) ;
\draw  [dash pattern={on 0.84pt off 2.51pt}]  (185.25,47) -- (198.89,89.39) ;
\draw  [draw opacity=0][fill={rgb, 255:red, 155; green, 155; blue, 155 }  ,fill opacity=0.25 ] (243.37,133.88) -- (167.25,210) -- (122.71,165.46) -- (198.84,89.34) -- cycle ;
\draw [color={rgb, 255:red, 0; green, 0; blue, 0 }  ,draw opacity=1 ]   (149.39,138.89) -- (122.71,165.46) ;
\draw [color={rgb, 255:red, 128; green, 128; blue, 128 }  ,draw opacity=1 ]   (122.71,165.46) -- (167.25,210) ;
\draw [color={rgb, 255:red, 128; green, 128; blue, 128 }  ,draw opacity=1 ]   (167.25,210) -- (243.37,133.88) ;

\draw (149,143) node [anchor=north] [inner sep=0.75pt]  [font=\scriptsize] [align=left] {$p$};
\draw (164.5,77.5) node [anchor=north west][inner sep=0.75pt]  [font=\scriptsize] [align=left] {${C\!H}^{+}$};
\draw (176.14,117.14) node [anchor=north west][inner sep=0.75pt]  [font=\scriptsize] [align=left] {$H^{+}$};
\draw (260.14,147.14) node [anchor=south west] [inner sep=0.75pt]  [font=\scriptsize] [align=left] {$\mathcal I^{+}$};
\draw (178.14,57.14) node [anchor=north west][inner sep=0.75pt]  [font=\scriptsize] [align=left] {$X$};
\draw (200.89,86.39) node [anchor=south west] [inner sep=0.75pt]  [font=\scriptsize] [align=left] {$i^{+}$};

\end{tikzpicture}

\caption{The extension through a Cauchy horizon for the Einstein--Maxwell-scalar field studied in \cite{Dafermos:2003vim,Dafermos:2003wr,Daf2005}. The light grey region extends to the event horizon $\mathcal H^+$ of the solution. The upper region is trapped and the evolution maximally extends to the Cauchy horizon ${CH}^+$. The metric is $C^0$ extendable across the Cauchy horizon in a region $X$. The results in \cite{Luk2019-I,Luk2019-II} exclude $C^2$ extendibility.}
\label{fig:daf2}
\end{figure}
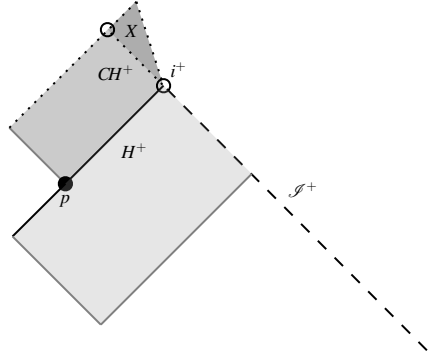

These results and the techniques employed have been constantly refined in order to tackle the stability problem of the Cauchy horizon in Reissner--Nordstrom  and later in Kerr spacetimes. This long programme has produced a huge literature in this sense and fostered a great interest on the study of stability properties of these features even outside the realm of spherical symmetry, see \cite{Daf2014} and more recently \cite{Kle2022} and again   \cite{Landsman:2022hrn} with therein references.  

\bigskip

As we have already said, Theorem \ref{thm:chr1} has been proved again in the paper \cite{An:2020vdf}, to carry on the argument in double null coordinates that are useful to study a generalization of the system that we have considered so far, the so called Einstein--Maxwell--\textit{charged} scalar field. 
In this situation the scalar field is coupled to the electromagnetic field, which results in a modification of Maxwell equations because Bianchi identity does not hold anymore separately for the two components of $T$.  We refer the reader to  \cite[Section 1.2.1]{VandeMoortel:2017ztd} of Maxime Van de Moortel for some mathematical and physical aspects of  this model. In the same paper,  the existence of a Cauchy horizon is proven, together with $C^0$--extendibility across this horizon. 
In \cite{An:2020vdf}, with suitable modifications of the sufficient conditions from the original theorem in \cite{Christodoulou:1991yfa}, the existence of trapped surface is proved for charged scalar field, confirming previous numerical insights on this model \cite{Torres:2014fga}. 

The above cited paper  \cite{VandeMoortel:2017ztd} actually also tackles the \textit{massive} case $m\ne 0$ (recall \eqref{eq:Lagr}), where Cauchy horizon existence is proven also where mass of the scalar field is taken into account.  In a later work 
\cite{VandeMoortel:2020olr} it is proven that the $C^2$--extension is not possible since Ricci curvature diverges.  Results about  continuous extendibility  have recently been found too  \cite{Kehle:2021jsp}.

\section{Scalar field collapse and the numerical approach}\label{sec:crit}

We have already commented in the Introduction that the ``numerical counterpart'' to Christodoulou for the scalar field collapse is represented by Choptuik and mostly by his celebrated \footnote{
Almost 1000 citations received 
at the time the present work was written, Scopus data.} paper \cite{Choptuik:1992jv} (but see also \cite{Chop92} for some more details on the numerical scheme). His work is important because it represents a crucial advance to understand the scalar field's dynamics, but also because it  revealed a behaviour that   was surprisingly found to be satisfied by many other relativistic models  along the years, as explained in the reviews  \cite{Gun2003,Gundlach:2007gc}.  Gundlach and his collaborators gave a huge contribution in understanding this surprising behaviour -- most of the material in the present section is taken from their works. 

Scalar field spherical model for numerical analysis purposes is  usually cast into a Schwarzschild gauge:
\begin{equation}
\label{eq:g-num}
-\alpha^2(t,r)\,\mathrm dt^2+a^2(t,r)\,\mathrm dr^2+ r^2 (\mathrm d\theta^2+\sin^2\theta\,\mathrm d\varphi^2). 
\end{equation}
Einstein equations \eqref{eq:EFE} in the free massless case become\footnote{It should be noted that there is no longer the normalization usually done in analytical studies to eliminate $\pi$ from Einstein field equations.}
\begin{subequations}
\begin{align}
\frac{a,_r}{a}+\frac{a^2-1}{2r}&=2\pi r(\Pi^2+\Phi^2),\label{eq:EFEn-1}\\
\frac{\alpha,_r}{\alpha}-\frac{a,_r}{a}-\frac{a^2-1}r&=0,\label{eq:EFEn-2}\\
\frac{a,_t}\alpha&=4\pi r\Phi\Pi,\label{eq:EFEn-3}
\end{align}
where 
$$
\Phi=\phi,_r,\qquad\Pi=\frac a\alpha \phi,_t.
$$
One advantage of using this gauge is that the first two equations contains only $r$--derivatives of the metric coefficients, and in principle they can be integrated in $r$ once the dynamics of the scalar field is known. The latter is captured by the wave equation $\Box\phi=0$ which takes the form
\begin{align}
\Phi,_t&=\left(\frac\alpha a\Pi\right),_r,\label{eq:Bianchin-1}\\
\Pi,_t&=\frac1{r^2}\left(r^2\frac{\alpha} a\Phi\right),_r.\label{eq:Bianchin-2}
\end{align}
\end{subequations}
Also notice that the mass in this formulation obey the relation
\begin{equation}
\label{eq:mass-chop}
m,_r=2\pi\left(\frac ra\right)^2(\Phi^2+\Pi^2),
\end{equation}
which allows quite simply to calculate the total mass of the spacetime at each time just by integration,
\begin{equation}
\label{eq:mass-t}
M(t)=2\pi\int_{0}^{\infty}\left(\frac ra\right)^2(\Phi^2+\Pi^2)\,\mathrm dr.
\end{equation}
The numerical evaluation is performed on a typical finite difference grid. $\Phi$ and $\Pi$ are specified at some initial time $t=0$, and this completely defines $\alpha(0,r)$ and $a(0,r)$. The numerical scheme ignores \eqref{eq:EFEn-3} and uses \eqref{eq:EFEn-1}--\eqref{eq:EFEn-2} together with \eqref{eq:Bianchin-1}--\eqref{eq:Bianchin-2} to determine completely the evolution of the system from the initial data -- for more details reader is referred to \cite{Chop92}, especially concerning how the discretization scale varies both in space and in time to keep the local truncation error below a fixed limit. 
Noticeably enough, the system is invariant under the rescaling
$$
(t,r)\to (kt, kr),\qquad\forall k>0;
$$
we'll get back on the symmetry features of this model in a while. 

\subsection{The black hole threshold}
To describe the nature of the critical behaviour found out by Choptuik,  let us
alternatively  assign $\phi_0(r)=\phi(0,r)$ instead of $\Phi(0,r)$ and $\Pi(0,r)$, for instance
\begin{equation}
\label{eq:phi0}
\phi_0(r)=\phi_0 r^3 e^{-\left(\frac{r-r_0}{\delta}\right)^q},
\end{equation}
 together with a condition that the scalar field is purely ingoing at initial time. Recalling the discussion in Section \ref{sec:trapped}, if the mass $2M\ll L$, where $L$ is the pulse thickness (controlled in \eqref{eq:phi0} by $\delta$) then the wave packet disperses, whereas when $2M\approx L$ then the strong gravity regime allows for trapped surfaces formation. What Choptuik did was to fix all the parameters of the initial data \eqref{eq:phi0} but one, call it $p$,  and vary it in order to obtain the critical value $p^*$ by a simple bisection technique -- that is, find two parameter values $p^{HI}$ and $p^{LO}$ such that $p>p^{HI}$ certainly  implies black hole and $p<p^{LO}$ certainly implies dispersion, and then proceed by subsequent bisections of the interval $[p^{LO},p^{HI}]$ to approximately determine  $p^*$. 

The study of this 1-parameter family of numerical solutions  revealed  that the mass of the black hole was obeying a scaling form depending on the free parameter $p$ when $p\to (p^*)^+$:
\begin{equation}
\label{eq:M-scale}
M\simeq C|p-p^*|^\gamma,
\end{equation}
where $C$ and, of course $p^*$, depend on the  parameter chosen to describe the family, but $\gamma$ is a universal constant ($\gamma\approx 0.37$) at least for this matter model. 

Moreover, the critical solution is \textit{discretely self--similar}, that is there exists a conformal isometry of the spacetime with conformal factor equal to $e^{-2\Delta}$ for some $\Delta$. This scale--periodicity (Christodoulou's terminology) or ``echoing'' (Choptuik's terminology) behaviour can be formalized as follows. Let us introduce the variable change 
$$
\tau=\log\frac{L}{t_*-t}, \qquad x=\log\frac r{t_*- t},
$$
then 
for values $t_*$ and $L$ depending on the model, 
the unknown functions $\alpha,\,a,\,\phi$ as a function of $(\tau,x)$ are periodic with respect to $\tau$ (for instance $\phi(\tau,x)=\phi(\tau+\Delta,x)$). For massless scalar field the period $\Delta$ is approximatively equal to $3.447$. 

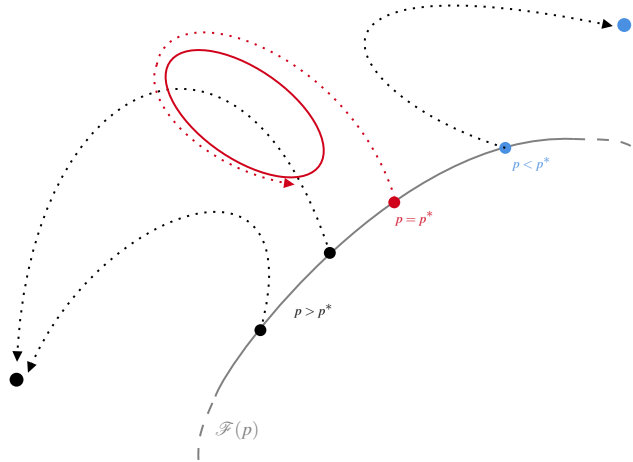
\begin{figure}
\centering

  
\tikzset {_j7zithvrw/.code = {\pgfsetadditionalshadetransform{ \pgftransformshift{\pgfpoint{0 bp } { 0 bp }  }  \pgftransformrotate{0 }  \pgftransformscale{2 }  }}}
\pgfdeclarehorizontalshading{_pq9aygwis}{150bp}{rgb(0bp)=(0.85,0.65,0.65);
rgb(37.5bp)=(0.85,0.65,0.65);
rgb(62.5bp)=(0.82,0.08,0.08);
rgb(100bp)=(0.82,0.08,0.08)}
\tikzset{_4u2bt5ent/.code = {\pgfsetadditionalshadetransform{\pgftransformshift{\pgfpoint{0 bp } { 0 bp }  }  \pgftransformrotate{0 }  \pgftransformscale{2 } }}}
\pgfdeclarehorizontalshading{_xd6lc4jqp} {150bp} {color(0bp)=(transparent!80);
color(37.5bp)=(transparent!80);
color(62.5bp)=(transparent!60);
color(100bp)=(transparent!60) } 
\pgfdeclarefading{_8orize5qj}{\tikz \fill[shading=_xd6lc4jqp,_4u2bt5ent] (0,0) rectangle (50bp,50bp); } 
\tikzset{every picture/.style={line width=0.75pt}} 

\begin{tikzpicture}[x=0.75pt,y=0.75pt,yscale=-1,xscale=1]

\draw  [color={rgb, 255:red, 0; green, 0; blue, 0 }  ,draw opacity=1 ][fill={rgb, 255:red, 0; green, 0; blue, 0 }  ,fill opacity=1 ] (104.5,246.5) .. controls (104.5,244.84) and (105.84,243.5) .. (107.5,243.5) .. controls (109.16,243.5) and (110.5,244.84) .. (110.5,246.5) .. controls (110.5,248.16) and (109.16,249.5) .. (107.5,249.5) .. controls (105.84,249.5) and (104.5,248.16) .. (104.5,246.5) -- cycle ;
\draw  [color={rgb, 255:red, 74; green, 144; blue, 226 }  ,draw opacity=1 ][fill={rgb, 255:red, 74; green, 144; blue, 226 }  ,fill opacity=1 ] (411,67.5) .. controls (411,65.84) and (412.34,64.5) .. (414,64.5) .. controls (415.66,64.5) and (417,65.84) .. (417,67.5) .. controls (417,69.16) and (415.66,70.5) .. (414,70.5) .. controls (412.34,70.5) and (411,69.16) .. (411,67.5) -- cycle ;
\draw  [draw opacity=0][shading=_pq9aygwis,_j7zithvrw,path fading= _8orize5qj ,fading transform={xshift=2}] (133.42,4.21) -- (132.12,147.44) -- (329.58,240.79) -- (330.88,97.56) -- cycle ;
\draw [color={rgb, 255:red, 128; green, 128; blue, 128 }  ,draw opacity=1 ]   (207.75,255) .. controls (215.75,233) and (301.25,121.5) .. (388.25,125) ;
\draw [color={rgb, 255:red, 128; green, 128; blue, 128 }  ,draw opacity=1 ] [dash pattern={on 4.5pt off 4.5pt}]  (199.25,287) .. controls (198.75,274.5) and (201.25,269.5) .. (207.75,255) ;
\draw [color={rgb, 255:red, 128; green, 128; blue, 128 }  ,draw opacity=1 ] [dash pattern={on 4.5pt off 4.5pt}]  (388.25,125) .. controls (401.75,125.5) and (409.75,124.5) .. (417.25,128.5) ;
\draw  [draw opacity=0][fill={rgb, 255:red, 208; green, 2; blue, 27 }  ,fill opacity=1 ] (295,157) .. controls (295,155.34) and (296.34,154) .. (298,154) .. controls (299.66,154) and (301,155.34) .. (301,157) .. controls (301,158.66) and (299.66,160) .. (298,160) .. controls (296.34,160) and (295,158.66) .. (295,157) -- cycle ;
\draw  [draw opacity=0][fill={rgb, 255:red, 0; green, 0; blue, 0 }  ,fill opacity=1 ] (227.5,221.53) .. controls (227.48,219.87) and (228.81,218.52) .. (230.47,218.5) .. controls (232.13,218.48) and (233.48,219.81) .. (233.5,221.47) .. controls (233.52,223.13) and (232.19,224.48) .. (230.53,224.5) .. controls (228.87,224.52) and (227.52,223.19) .. (227.5,221.53) -- cycle ;
\draw  [draw opacity=0][fill={rgb, 255:red, 74; green, 144; blue, 226 }  ,fill opacity=1 ] (351,129.5) .. controls (351,127.84) and (352.34,126.5) .. (354,126.5) .. controls (355.66,126.5) and (357,127.84) .. (357,129.5) .. controls (357,131.16) and (355.66,132.5) .. (354,132.5) .. controls (352.34,132.5) and (351,131.16) .. (351,129.5) -- cycle ;
\draw  [dash pattern={on 0.84pt off 2.51pt}]  (354,129.5) .. controls (268.68,103.13) and (233.11,31.72) .. (405.63,67.45) ;
\draw [shift={(408.25,68)}, rotate = 191.91] [fill={rgb, 255:red, 0; green, 0; blue, 0 }  ][line width=0.08]  [draw opacity=0] (5.36,-2.57) -- (0,0) -- (5.36,2.57) -- cycle    ;
\draw  [dash pattern={on 0.84pt off 2.51pt}]  (230.5,221.5) .. controls (257.97,122) and (148.97,159.72) .. (114.28,240.54) ;
\draw [shift={(113.25,243)}, rotate = 292.1] [fill={rgb, 255:red, 0; green, 0; blue, 0 }  ][line width=0.08]  [draw opacity=0] (5.36,-2.57) -- (0,0) -- (5.36,2.57) -- cycle    ;
\draw  [draw opacity=0][fill={rgb, 255:red, 0; green, 0; blue, 0 }  ,fill opacity=1 ] (262.5,182.5) .. controls (262.5,180.84) and (263.84,179.5) .. (265.5,179.5) .. controls (267.16,179.5) and (268.5,180.84) .. (268.5,182.5) .. controls (268.5,184.16) and (267.16,185.5) .. (265.5,185.5) .. controls (263.84,185.5) and (262.5,184.16) .. (262.5,182.5) -- cycle ;
\draw  [dash pattern={on 0.84pt off 2.51pt}]  (265.5,182.5) .. controls (220.48,43.2) and (114.32,92.5) .. (107.84,235.34) ;
\draw [shift={(107.75,237.5)}, rotate = 272.18] [fill={rgb, 255:red, 0; green, 0; blue, 0 }  ][line width=0.08]  [draw opacity=0] (5.36,-2.57) -- (0,0) -- (5.36,2.57) -- cycle    ;
\draw  [color={rgb, 255:red, 208; green, 2; blue, 27 }  ,draw opacity=1 ] (257.9,116.72) .. controls (268.09,134.29) and (260.54,146.54) .. (241.02,144.07) .. controls (221.51,141.6) and (197.42,125.35) .. (187.22,107.78) .. controls (177.03,90.21) and (184.59,77.96) .. (204.1,80.43) .. controls (223.62,82.9) and (247.7,99.15) .. (257.9,116.72) -- cycle ;
\draw [color={rgb, 255:red, 208; green, 2; blue, 27 }  ,draw opacity=1 ] [dash pattern={on 0.84pt off 2.51pt}]  (178.25,87.5) .. controls (185.25,49.5) and (274.25,80.5) .. (298,154) ;
\draw [color={rgb, 255:red, 208; green, 2; blue, 27 }  ,draw opacity=1 ] [dash pattern={on 0.84pt off 2.51pt}]  (178.25,87.5) .. controls (169.92,106.61) and (201.93,140.13) .. (245.09,147.58) ;
\draw [shift={(247.75,148)}, rotate = 188.31] [fill={rgb, 255:red, 208; green, 2; blue, 27 }  ,fill opacity=1 ][line width=0.08]  [draw opacity=0] (5.36,-2.57) -- (0,0) -- (5.36,2.57) -- cycle    ;

\draw (206.25,266) node [anchor=north west][inner sep=0.75pt]  [font=\scriptsize,color={rgb, 255:red, 128; green, 128; blue, 128 }  ,opacity=1 ] [align=left] {$\displaystyle \mathcal{F}( p)$};
\draw (246,206.5) node [anchor=north west][inner sep=0.75pt]  [font=\tiny,color={rgb, 255:red, 0; green, 0; blue, 0 }  ,opacity=1 ] [align=left] {$\displaystyle p >p^{*}$};
\draw (356,132.5) node [anchor=north west][inner sep=0.75pt]  [font=\tiny,color={rgb, 255:red, 74; green, 144; blue, 226 }  ,opacity=1 ] [align=left] {$\displaystyle p< p^{*}$};
\draw (297,160) node [anchor=north west][inner sep=0.75pt]  [font=\tiny,color={rgb, 255:red, 208; green, 2; blue, 27 }  ,opacity=1 ] [align=left] {$\displaystyle p=p^{*}$};

\end{tikzpicture}

\caption{The schematic picture of the critical behaviour \cite[Figure 2]{Gun2003}. Points in the space represent initial data for the scalar collapse evolution, dotted lines represent the dynamical evolution of the data. The red surface is the black hole threshold. Every one parameter family $\mathcal F(p)$ of initial data (grey curve) intersects the black hole threshold for a value of the parameter $p=p^*$ (the red point on $\mathcal F(p)$) that evolves near the critical solution, represented by the red limit cycle because of its periodicity property.  When $p<p^*$ (blue points) then the spacetime evolves to a flat configuration where scalar field is dispersed at infinity. When $p>p^*$, then trapped surface forms and a black hole forms.}
\label{fig:threshold}
\end{figure}

Therefore, representing the space of initial data as the phase space associated to the dynamics of scalar field collapse, the situation is roughly explained in Fig. \ref{fig:threshold}.  The stable manifold for the critical solution is the so--called\textit{ black hole threshold}, and separates data evolving to a blackhole to those that dispersing to future infinity. This is a rough description for many reasons: one of these is that the unstable mode for the black hole is not unique, as pointed out in \cite{Frolov:1999fv}. Moreover, it's the numerical analysis made by  considering several 1-parameter families of initial data, all exhibiting the same behaviour, that supports this universality, towards the conclusion that the black hole threshold is a codimension one submanifold.
 
It is to remember that the scalar field case is the first example exhibiting features of \textit{universality}, -- solutions' endstates depend on the positions of their data with respect to a critical submanifold only -- \textit{black hole mass scaling} like \eqref{eq:M-scale},  and critical solution's \textit{scale--invariance} of some kind.  Attempts to study the critical behaviour have been made \cite{Bhattacharya:2010tc,Zhang:2014dfa} with both quantitative and qualitative results not completely overlapping with those from numerical studies. All these hints are  probably the signature of profound phenomena that have yet to be entirely understood in their full generality. 

What's for sure, is that numerical analysis has confirmed critical behaviour existence in many cases. In the realm of scalar field collapse, when considering a massive scalar field \cite{Brady:1997fj}, in the non-minimally coupled case with a potential \cite{Jimenez-Vazquez:2021dns}, and in other gravity theories like Brans--Dicke theory \cite{Choptuik:1997rt}. The latter case is relevant since in this context a proof of the existence of naked singularity, extending to this case the example of Christodoulou \cite{Christodoulou:1994hg} that we have already cited and that we will discuss in next Subsection too, is given \cite{Bedjaoui:2010nh}. Further examples of critical behaviour in other matter models are given for instance in the case of the Einstein cluster studied in \cite{Harada:2007mq,Mahajan:2007vw}
For a much broader overview on critical behaviour we  refer the reader to \cite{Gundlach:2007gc} once again and to the more recent review \cite{Chop2015} also. 

\subsection{Naked singularity in the critical case}\label{sec:crit-ns}

The  picture sketched in the above subsection  shows that there is a so--called \textit{critical solution} that lives in the threshold black hole and is neither regular nor contains a  black hole covered by a horizon. In Fig. \ref{fig:threshold} this solution is represented as a limit cycle in the phase space since, as we have observed, it has periodicity features. 
This critical solution, as mentioned in the previous sections, is important also because it  possesses a singularity that yet is not covered by a event horizon, and is visible a future null infinity. It is therefore a naked singularity. 

The global structure of this spacetime has been the object of several works \cite{Gundlach:1995kd,Gundlach:1996eg,Gundlach:2003pg,Martin-Garcia:2003xgm}. However it must be remembered that its existence basically comes from numerical inference, although quite recently a computer--assisted proof appeared \cite{Reiterer:2012hnr} that, starting from and approximate solution shows the existence of a real analytic \textit{true} solution, precisely interpreted as the Choptuik critical case.

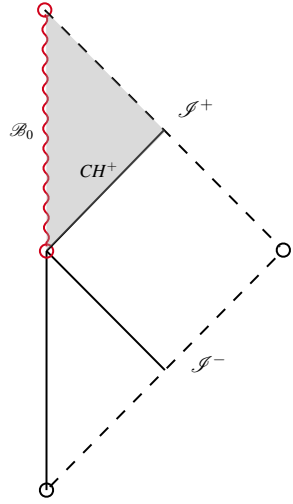
\begin{figure}
\centering

\tikzset{every picture/.style={line width=0.75pt}} 

\begin{tikzpicture}[x=0.75pt,y=0.75pt,yscale=-1,xscale=1]

\draw    (119.83,169.67) -- (119.8,290.35) ;
\draw  [dash pattern={on 4.5pt off 4.5pt}]  (121.45,288.67) -- (239.25,169.21) ;
\draw [shift={(119.8,290.35)}, rotate = 314.6] [color={rgb, 255:red, 0; green, 0; blue, 0 }  ][line width=0.75]      (0, 0) circle [x radius= 3.35, y radius= 3.35]   ;
\draw    (119.83,169.67) -- (179.52,229.78) ;
\draw [color={rgb, 255:red, 208; green, 2; blue, 27 }  ,draw opacity=1 ]   (118.92,50.35) .. controls (120.6,52) and (120.61,53.67) .. (118.96,55.35) .. controls (117.31,57.03) and (117.32,58.7) .. (119,60.35) .. controls (120.68,62) and (120.69,63.67) .. (119.04,65.35) .. controls (117.39,67.03) and (117.4,68.7) .. (119.08,70.35) .. controls (120.75,72.01) and (120.76,73.68) .. (119.11,75.35) .. controls (117.46,77.03) and (117.47,78.7) .. (119.15,80.35) .. controls (120.83,82) and (120.84,83.67) .. (119.19,85.35) .. controls (117.54,87.03) and (117.55,88.7) .. (119.23,90.35) .. controls (120.91,92) and (120.92,93.67) .. (119.27,95.35) .. controls (117.62,97.02) and (117.63,98.69) .. (119.3,100.35) .. controls (120.98,102) and (120.99,103.67) .. (119.34,105.35) .. controls (117.69,107.03) and (117.7,108.7) .. (119.38,110.35) .. controls (121.06,112) and (121.07,113.67) .. (119.42,115.35) .. controls (117.77,117.03) and (117.78,118.7) .. (119.46,120.35) .. controls (121.13,122.01) and (121.14,123.68) .. (119.49,125.35) .. controls (117.84,127.03) and (117.85,128.7) .. (119.53,130.35) .. controls (121.21,132) and (121.22,133.67) .. (119.57,135.35) .. controls (117.92,137.03) and (117.93,138.7) .. (119.61,140.35) .. controls (121.29,142) and (121.3,143.67) .. (119.65,145.35) .. controls (118,147.02) and (118.01,148.69) .. (119.68,150.35) .. controls (121.36,152) and (121.37,153.67) .. (119.72,155.35) .. controls (118.07,157.03) and (118.08,158.7) .. (119.76,160.35) .. controls (121.44,162) and (121.45,163.67) .. (119.8,165.35) -- (119.81,167.32) -- (119.81,167.32) ;
\draw [shift={(119.83,169.67)}, rotate = 89.56] [color={rgb, 255:red, 208; green, 2; blue, 27 }  ,draw opacity=1 ][line width=0.75]      (0, 0) circle [x radius= 3.35, y radius= 3.35]   ;
\draw [shift={(118.91,48)}, rotate = 89.56] [color={rgb, 255:red, 208; green, 2; blue, 27 }  ,draw opacity=1 ][line width=0.75]      (0, 0) circle [x radius= 3.35, y radius= 3.35]   ;
\draw  [dash pattern={on 4.5pt off 4.5pt}]  (118.91,48) -- (237.59,167.54) ;
\draw [shift={(239.25,169.21)}, rotate = 45.21] [color={rgb, 255:red, 0; green, 0; blue, 0 }  ][line width=0.75]      (0, 0) circle [x radius= 3.35, y radius= 3.35]   ;
\draw    (119.83,169.67) -- (179.08,108.61) ;
\draw  [draw opacity=0][fill={rgb, 255:red, 155; green, 155; blue, 155 }  ,fill opacity=0.37 ] (119.6,48.71) -- (119.83,169.68) -- (179.78,109.31) -- cycle ;

\draw (99.71,103.65) node [anchor=north west][inner sep=0.75pt]  [font=\scriptsize] [align=left] {$\displaystyle \mathcal{B}_{0}$};
\draw (184.4,91.37) node [anchor=north west][inner sep=0.75pt]   [align=left] {$\mathcal I^{+}$};
\draw (190.89,220.26) node [anchor=north west][inner sep=0.75pt]   [align=left] {$\mathcal I^{-}$};
\draw (135.44,123.32) node [anchor=north west][inner sep=0.75pt]  [font=\scriptsize] [align=left] {$\displaystyle CH^{+}$};

\end{tikzpicture}

\caption{The critical solution with a possible $C^0$ extension \cite[Figure 2]{Martin-Garcia:2003xgm}. The grey region represent a possible $C^0$ extension of the maximal analytical extension of Choptuik critical spacetime across the Cauchy horizon ${CH}^+$. In the maximal analytical extension, the naked central singularity is the lower end of the red vertical line, where the regular centre (black vertical line) terminates. }
\label{fig:gund1}
\end{figure}

The picture emerging from the analysis of Gundlach and Mart\'{i}n--Garc\'{i}a (see in particular  \cite{Martin-Garcia:2003xgm}) shows that it is a spacetime that can be uniquely continued up to the future outgoing null cone of the pointlike singularity, that therefore is a Cauchy horizon. This analysis seems confirmed also using other numerical approaches, see e.g. \cite{Frolov:2003dk}.

The solution can be extended in a non unique way beyond the Cauchy horizon, but the only discrete self--similar possible extensions are either with a regular centre or with a timelike central singularity (see Figure \ref{fig:gund1}), excluding null singularity followed by a spacelike dynamical singularity, similar to other collapsing matter model such as spherical dust. The past light cone is a self similar horizon -- a null surface invariant by the self--similarity. 

Without considering the possible extension across the Cauchy horizon, the model has then a pointlike singularity, in that resembling Christodoulou counterexample built in \cite{Christodoulou:1994hg}. The two solutions anyway cannot coincide, because the latter  is actually a \textit{continuous} self--similar solution, with metric
\begin{equation}
\label{eq:chr-ns}
 g=-e^{2\nu}\,\mathrm du^2-2e^{\lambda+\nu}\,\mathrm du\,\mathrm dr+r^2\gamma_{\mathbb S^2},
\end{equation}
where $\lambda$ and $\nu$ are function of
$$
x=-\frac{r}{u} 
$$
In this way, defined the vector field $S=u\partial_u+r\partial_r$, one has $\mathcal L_S g=2g$, $Sr=r$. Moreover the vector field acts on the scalar field $\phi$ so that
$S\phi=-k\in\mathbb R$.

Self--similarity reduces Einstein equations to an ODE system, and in \cite{Christodoulou:1994hg} it is shown that a solution without a horizon exists in the half--plane $u<0$, such that the scalar field along two null directions terminating in $(u=0,r=0)$ diverges like $\log r$, and this makes $(0,0)$ a singular point not covered by a horizon. In this example the solution is extended to the half plane $u>0$ isometrically, so that one can obtain a solution that is regular everywhere except one single point. 

At this stage, one may argue that Christodoulou analysis described in Section \ref{sec:trapped} and the picture emerging from critical behaviour  don't overlap, since one may apply the instability argument described in Section \ref{sec:instability} above to an initial data in the black hole threshold, obtaining a 2-parameter family of initial data always leading to black hole formation. As pointed out in \cite{Gun2003}, one possible explanation for this apparent contradiction lies in that the picture emerging from numerical evidence concerns solutions starting from \textit{smooth} functions, unlike the argument presented in Subsection \ref{sec:instability}. 

\begin{figure}
\centering

\tikzset{every picture/.style={line width=0.75pt}} 

\begin{tikzpicture}[x=0.75pt,y=0.75pt,yscale=-1,xscale=1]

\draw    (60.41,112.31) -- (60.25,268.65) ;
\draw [shift={(60.25,271)}, rotate = 90.06] [color={rgb, 255:red, 0; green, 0; blue, 0 }  ][line width=0.75]      (0, 0) circle [x radius= 3.35, y radius= 3.35]   ;
\draw [shift={(60.42,109.96)}, rotate = 90.06] [color={rgb, 255:red, 0; green, 0; blue, 0 }  ][line width=0.75]      (0, 0) circle [x radius= 3.35, y radius= 3.35]   ;
\draw  [dash pattern={on 4.5pt off 4.5pt}]  (140.25,189.98) -- (60.25,271) ;
\draw [color={rgb, 255:red, 74; green, 144; blue, 226 }  ,draw opacity=1 ] [dash pattern={on 4.5pt off 4.5pt}]  (218.58,112.61) -- (140.25,189.98) ;
\draw [shift={(220.25,110.96)}, rotate = 135.35] [color={rgb, 255:red, 74; green, 144; blue, 226 }  ,draw opacity=1 ][line width=0.75]      (0, 0) circle [x radius= 3.35, y radius= 3.35]   ;
\draw  [dash pattern={on 4.5pt off 4.5pt}]  (190.75,81.96) -- (220.25,110.96) ;
\draw  [color={rgb, 255:red, 155; green, 155; blue, 155 }  ,draw opacity=1 ][fill={rgb, 255:red, 155; green, 155; blue, 155 }  ,fill opacity=0.36 ] (130.25,153.46) -- (153.51,176.72) -- (140.25,189.98) -- (116.99,166.72) -- cycle ;
\draw [color={rgb, 255:red, 74; green, 144; blue, 226 }  ,draw opacity=1 ]   (60.42,109.96) -- (140.25,189.98) ;
\draw [color={rgb, 255:red, 74; green, 144; blue, 226 }  ,draw opacity=1 ]   (140.25,189.98) -- (153.51,176.72) ;
\draw [color={rgb, 255:red, 208; green, 2; blue, 27 }  ,draw opacity=1 ][fill={rgb, 255:red, 208; green, 2; blue, 27 }  ,fill opacity=0.5 ]   (124.03,159.85) .. controls (129.03,163.85) and (135.53,159.35) .. (135.53,159.35) ;
\draw  [draw opacity=0][fill={rgb, 255:red, 208; green, 2; blue, 27 }  ,fill opacity=0.5 ] (135.53,159.35) -- (124.03,159.85) -- (130.09,153.85) -- cycle ;
\draw [color={rgb, 255:red, 128; green, 128; blue, 128 }  ,draw opacity=1 ]   (183,259.67) -- (157.63,229.86) ;
\draw [shift={(156.33,228.33)}, rotate = 49.6] [color={rgb, 255:red, 128; green, 128; blue, 128 }  ,draw opacity=1 ][line width=0.75]    (10.93,-3.29) .. controls (6.95,-1.4) and (3.31,-0.3) .. (0,0) .. controls (3.31,0.3) and (6.95,1.4) .. (10.93,3.29)   ;
\draw [color={rgb, 255:red, 128; green, 128; blue, 128 }  ,draw opacity=1 ]   (183,259.67) -- (212.16,234.31) ;
\draw [shift={(213.67,233)}, rotate = 138.99] [color={rgb, 255:red, 128; green, 128; blue, 128 }  ,draw opacity=1 ][line width=0.75]    (10.93,-3.29) .. controls (6.95,-1.4) and (3.31,-0.3) .. (0,0) .. controls (3.31,0.3) and (6.95,1.4) .. (10.93,3.29)   ;

\draw (183.5,165) node [anchor=north west][inner sep=0.75pt]   [align=left] {$\mathcal I^{-}$};
\draw (154.33,231.33) node [anchor=north east] [inner sep=0.75pt]  [font=\scriptsize,color={rgb, 255:red, 128; green, 128; blue, 128 }  ,opacity=1 ] [align=left] {$\displaystyle u$};
\draw (215.67,236) node [anchor=north west][inner sep=0.75pt]  [font=\scriptsize,color={rgb, 255:red, 128; green, 128; blue, 128 }  ,opacity=1 ] [align=left] {$\displaystyle v$};
\draw (142.25,192.98) node [anchor=north west][inner sep=0.75pt]  [font=\scriptsize] [align=left] {$\displaystyle ( u_{0} ,0)$};
\draw (35,99) node [anchor=north west][inner sep=0.75pt]  [font=\footnotesize] [align=left] {$\displaystyle \mathcal{B}_{0}$};
\draw (84,224.01) node [anchor=north west][inner sep=0.75pt]  [font=\scriptsize,rotate=-315] [align=left] {$\displaystyle v< 0$};
\draw (161,148.01) node [anchor=north west][inner sep=0.75pt]  [font=\scriptsize,color={rgb, 255:red, 74; green, 144; blue, 226 }  ,opacity=1 ,rotate=-315] [align=left] {$\displaystyle v >0$};

\end{tikzpicture}

\caption{Non-smoothness of data in Christodoulou instability argument. The argument is developed in the exterior of the past null cone of $\mathcal B_0$, defining suitable data for $v>0$ (blue dashed line) in order to determine an apparent horizon (in the picture only the first element of the sequence of trapped surfaces is shown, see also Fig. \ref{fig:chr4}). The data are then \textit{non-smoothly} extended  for $v<0$ (black dashed line).}
\label{fig:data}
\end{figure}
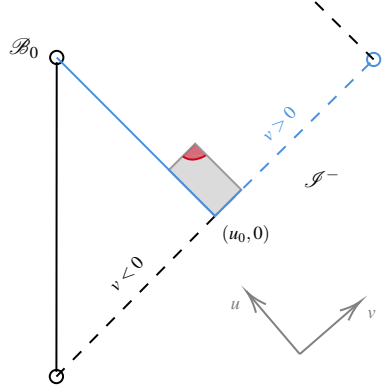

Indeed (see Fig. \ref{fig:data}) let us consider  initial data $\alpha_0$ corresponding to a solution in the black hole threshold. In view of the above discussion we can suppose this data at past null infinity $I^-$, in order to be consistent with the setup of Section \ref{sec:PDE}. The perturbative argument of Christodoulou considers 
two functions $f_1(v)$ and $f_2(v)$ (see step 3. in Section \ref{sec:instability}). These functions will be  set to zero for $v<0$, so that the evolution is left unchanged in the interior of the past null cone of $\mathcal B_0$. Then they are defined for $v>0$ in order to obtain an apparent horizon extendable until $\mathcal B_0$, as explained in Section \ref{sec:instability}. This results in that $f_1(v)$ has a jump discontinuity at $v=0$, while $f_2(v)$ is absolutely continuous on the real line, but not smooth. 

Therefore, although 
the whole 2-parameters family of initial data
$$\mathcal F(p_1,p_2)=\{\alpha_0(v)+p_1 f_1(v)+p_2 f_2(v)\,:\,p_1,p_2\in\mathcal R\}
$$
evolves to a black hole, this is not a family of smooth initial data, and then lies outside the range covered by numerical studies.

\section{Homogeneous scalar fields}\label{sec:hsf}

The scalar field collapsing model that we have discussed so far has no constraint imposed on the evolution of the scalar field, excluding the initial assumption that we have made at the very beginning that $\phi$ is defined on the quotient space $\mathcal Q$ for symmetry reasons. In a cosmological setting, however, isotropy arguments impose the further constraint that $\phi$ is a function on the cosmological time only. Of course this considerably reduce the mathematical complexity of the model, that will be now ruled by a system of ODEs. On the other side, this allows to include without significant drawbacks the self--interaction potential $V(\phi)$, and to study solutions' evolution by means of dynamical systems techniques \cite{Gunzig:2000ce}, retaining information on the qualitative behaviour of the scalar field depending on assumptions made on $V(\phi)$. 

The physical significance of scalar fields with potential in a cosmological setting is related to the interpretation of inflationary expansion phase of the universe \cite{Goldwirth:1991rj}. But we find them also in string theory or in higher dimensional gravity, using Kaluza--Klein reduction \cite{Zhang:2015rsa}. Moreover, they arise also in nonlinear theories of gravitation \cite{Magnano:1987zz,Ferraris:1988zz} such as $f(R)$ theories, since as is well known they can be reduced, up to a conformal transformation, to the ordinary gravity theory where the energy momentum tensor has a scalar field component, where $V(\phi)$ depends on the form of $f(R)$. Scalar field and spacetime singularities are also connected of course within the study of primordial gravitation, for instance to suggest alternative model to the
standard $\Lambda CDM$ paradigm  \cite{Luongo:2018lgy} or within loop quantum cosmology \cite{Ashtekar:2011ni}.

Perhaps the best known example of homogeneous collapsing solutions is the Oppenheimer--Snyder \cite{Oppenheimer:1939ue} homogeneous dust ball collapsing to a black hole. Similarly, one may consider a scalar field -- possibly coupled with other matter types -- as interior of a star, that must be matched with a reasonable exterior to get a global model where one can, once again, study the collapse evolution and see whether singularities form with or without a horizon. In this context one can build homogeneous ``scalar field stars'' matching a FR interior with a -- possibly non homogeneous -- exterior solution. 

All in all, there is a huge variety of situations that have been studied, and a wide account on the existing literature on the subject can be found in  the  two companion reviews 
\cite{Leon:2020pfy,Leon:2020ovw}.

Einstein's field equations for the RW model
\begin{equation}\label{eq:FRW}
\mathrm ds^2=-dt^2 + a(t)^2\left[\frac1{1-k r^2}\,\mathrm
dr^2+r^2\left(\mathrm d\theta^2+\sin^2\theta\,\mathrm
d\varphi^2\right)\right],
\end{equation}
with the stress--energy tensor \eqref{eq:T}
are 
\begin{align}
&(G^0_0=8\pi T^0_0):&\qquad&-\frac{3(k+\dot
a^2)}{a^2}=-\left(\frac12\dot\phi^2+V(\phi)\right),
\label{eq:G00}\\
&(G^1_1=8\pi T^1_1):&\qquad&-\frac{(k+\dot a^2)+2a\ddot
a}{a^2}=\left(\frac12\dot\phi^2-V(\phi)\right).\label{eq:G11}
\end{align}
in the unknown functions $a(t),\phi(t)$.
\footnote{As in the previous situations, the scalar field and the potential are normalized in order to get rid of the $\pi$ factor in  Einstein field equations. Unfortunately, this normalization changes from paper to paper in literature -- here we will follow the normalization made (for the flat $k=0$ case) in \cite{Giambo:2008ya}, that will possibly differ from those used in other papers that we will refer to.}
These equations imply as usual Bianchi identity that in this context becomes
\begin{equation}\label{eq:KG}
T^\mu_{\,\,0;\mu}=-\dot\phi\left(\ddot\phi+V'(\phi)+3\frac{\dot
a}a\dot\phi\right)=0.
\end{equation}
The above second order system
can be translated into a first order system introducing
the \textit{Hubble function }
$$
h=\frac{\dot a}a,
$$ and $v=\dot\phi$. 
We first briefly consider solutions such that 
 $\phi(t)=\phi_0$ constant on some
interval so \eqref{eq:KG} holds, but $V'(\phi_0)\ne 0$ (so the term in round bracket in \eqref{eq:KG} is non zero). These evolutions are obtained by \eqref{eq:G11}:
$$
\dot a^2=\frac{V(\phi_0)}3 a^2-k, 
$$
corresponding to a (anti)deSitter solution (the regular value of the potential in $\phi_0$ plays the role of the cosmological constant). Excluding this case we have that $(\phi,v,h)$ solves the regular system
\begin{subequations}
\begin{align}
\dot\phi&=v,\label{eq:dphi}\\
\dot v&=-V'(\phi)-3h\,v,\,\label{eq:dv}\\
\dot h&=-h^2-\tfrac13(v^2-V(\phi)).\label{eq:dh}
\end{align}
\end{subequations}
Observe that the function
\begin{equation}\label{eq:W}
W(\phi,v,h)=h^2-\frac13\left(\frac12v^2+V(\phi)\right)
\end{equation}
satisfies by \eqref{eq:G00}  
 the identity
\begin{equation}\label{eq:Wk}
W(t)=-\frac{k}{a(t)^2},
\end{equation}
where $k=-\mathrm{sgn}(W(0))$. The sign of $W$ is  invariant
by the flow ($\dot W=-2hW$) -- in other words the system \eqref{eq:dphi}--\eqref{eq:dh} represents at once \textit{all} RW cosmologies \eqref{eq:FRW}. 

The equilibrium points are given by
$(\phi_*,0,h_*)$ such that $V'(\phi_*)=0,\,h_*^2=\frac13 V(\phi_*)$
(notice that $W(\phi_*,0,h_*)=0$), that
physically correspond to (anti)deSitter universe.
The sign of the Hubble function $h$ tells us whether the solution is expanding ($h>0$) or collapsing ($h<0$).

\subsection{(Re)collapsing solutions and nature of the singularity}
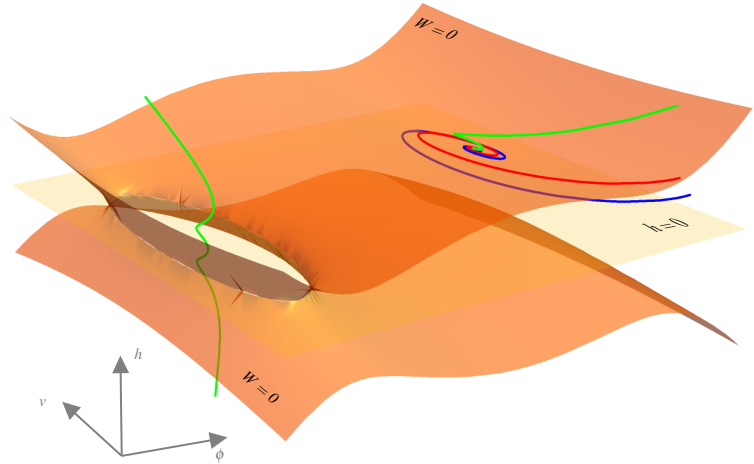
\begin{figure}
\centering

\tikzset{every picture/.style={line width=0.75pt}} 

\begin{tikzpicture}[x=0.75pt,y=0.75pt,yscale=-1,xscale=1]

\draw (232.08,174.83) node  {\includegraphics[width=293.88pt,height=197.25pt]{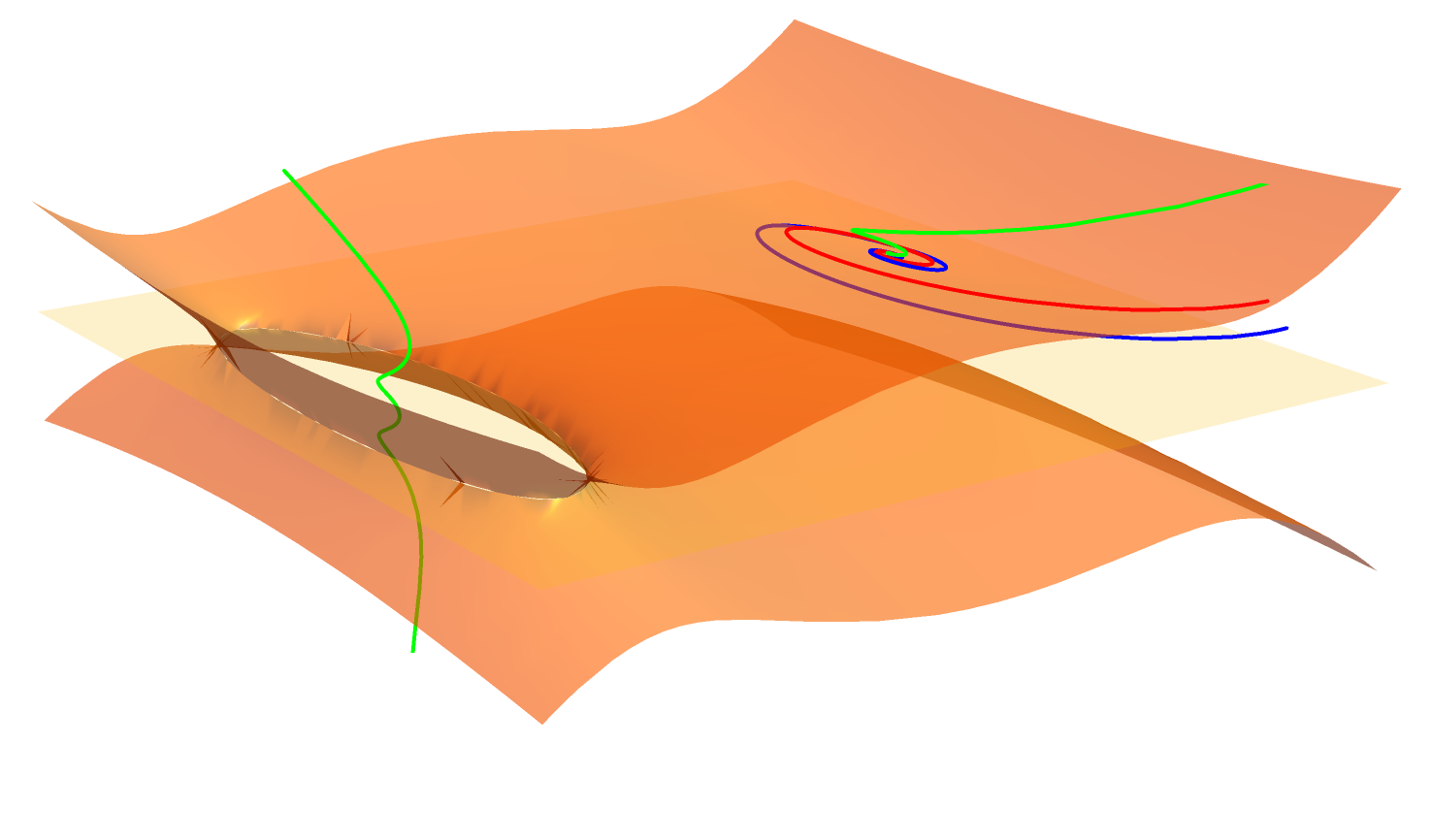}};
\draw [color={rgb, 255:red, 128; green, 128; blue, 128 }  ,draw opacity=1 ]   (102,277.67) -- (101.37,230) ;
\draw [shift={(101.33,227)}, rotate = 89.25] [fill={rgb, 255:red, 128; green, 128; blue, 128 }  ,fill opacity=1 ][line width=0.08]  [draw opacity=0] (8.93,-4.29) -- (0,0) -- (8.93,4.29) -- cycle    ;
\draw [color={rgb, 255:red, 128; green, 128; blue, 128 }  ,draw opacity=1 ]   (74.22,252.35) -- (102,277.67) ;
\draw [shift={(72,250.33)}, rotate = 42.34] [fill={rgb, 255:red, 128; green, 128; blue, 128 }  ,fill opacity=1 ][line width=0.08]  [draw opacity=0] (8.93,-4.29) -- (0,0) -- (8.93,4.29) -- cycle    ;
\draw [color={rgb, 255:red, 128; green, 128; blue, 128 }  ,draw opacity=1 ]   (102,277.67) -- (152.38,268.85) ;
\draw [shift={(155.33,268.33)}, rotate = 170.07] [fill={rgb, 255:red, 128; green, 128; blue, 128 }  ,fill opacity=1 ][line width=0.08]  [draw opacity=0] (8.93,-4.29) -- (0,0) -- (8.93,4.29) -- cycle    ;

\draw (360.38,160.23) node [anchor=north west][inner sep=0.75pt]  [font=\footnotesize,rotate=-349.27,xslant=-1.45,xscale=0.7,yscale=0.7] [align=left] {$\displaystyle h=0$};
\draw (253.13,55.24) node [anchor=north west][inner sep=0.75pt]  [font=\footnotesize,rotate=-20.7,xslant=0.54,xscale=0.7,yscale=0.7] [align=left] {$\displaystyle W=0$};
\draw (166.25,233.39) node [anchor=north west][inner sep=0.75pt]  [font=\footnotesize,rotate=-39.75,xslant=0.16,xscale=0.7,yscale=0.7] [align=left] {$\displaystyle W=0$};
\draw (148,272.67) node [anchor=north west][inner sep=0.75pt]  [font=\footnotesize,color={rgb, 255:red, 128; green, 128; blue, 128 }  ,opacity=1 ,xscale=0.7,yscale=0.7] [align=left] {$\displaystyle \phi $};
\draw (59.33,248) node [anchor=north west][inner sep=0.75pt]  [font=\footnotesize,color={rgb, 255:red, 128; green, 128; blue, 128 }  ,opacity=1 ,xscale=0.7,yscale=0.7] [align=left] {$\displaystyle v$};
\draw (107.33,222) node [anchor=north west][inner sep=0.75pt]  [font=\footnotesize,color={rgb, 255:red, 128; green, 128; blue, 128 }  ,opacity=1 ,xscale=0.7,yscale=0.7] [align=left] {$\displaystyle h$};

\end{tikzpicture}

\caption{The phase space of system \eqref{eq:dphi}--\eqref{eq:dh}, in a case where $V(\phi)$ has two local minima, one with negative critical value, the other with positive critical value (see \cite[Fig. 1]{Giambo:2008ck}). Example of cosmologies for the cases $k=-1$ (green curves), $k=0$ (red curve) and $k=1$ (blue curve) starting in the half space $h>0$ (expanding initial conditions) are sketched.}
\label{fig:homo1}
\end{figure}

Let us first briefly discuss the evolution from expanding initial conditions, i.e. $h(0)>0$  \cite{Miritzis:2003ym}.  Fig. \ref{fig:homo1} sketches some of the integral curves of system \eqref{eq:dphi}--\eqref{eq:dh} in this case. A minimum  of the potential $V(\phi)$  with positive critical value is a stable equilibrium point, and nearby cosmologies approach the corresponding deSitter state asymptotically: no collapse will take place here. These are the curves on the right of Fig. \ref{fig:homo1}.

For the open topologies $k\le 0$, corresponding to the portion of the phace space $W\ge 0$, the  upper branch of $W=0$ near this local minimum acts like a constraint forcing the solutions to approach the minimum. The deSitter space is attractive also for the portion ($W<0$) of the phase space between the two branches of $W=0$, corresponding to  $k=1$ cosmologies, at least when the minimum of $V(\phi)$ is \textit{strictly} positive. When $V(\phi_0)=0$ then recollapse is possible in $k=1$ case. 

On the left of Fig. \ref{fig:homo1} we have a situation near a local minimum for $V(\phi)$ with negative critical value. In this case the two branches of $W>0$ intersect somewhere at $h=0$, determining a sort of ``tunnel'' crossed by solutions in the open topology $k=-1$, that in this way reach the $h<0$ portion of the space and then recollapse. A fortiori, that happens also for solutions with $k=0$ and $k=1$.

Once we have established that homogeneous scalar fields may collapse, even if they start expanding, we can ask whether the evolution ends into a singularity and if a horizon forms. Since the dependence on $h$ in the right-hand side of \eqref{eq:dh} is quadratic, a singularity will likely form in a finite amount of cosmological time. In \cite{Giambo:2005se} the flat case $k=0$ is approached by using the scale factor $a(t)$ as a sort of time, and expressing the energy density $\rho_{SF}$ of the model, given by (minus) the right-hand side in \eqref{eq:G00}, as
\begin{equation}
\label{eq:eps}
\rho_{SF}(t):=\frac12\dot\phi(t)+V(\phi(t))=3\left(\frac{\psi(a(t))}{a(t)}\right)^2.
\end{equation}
Using \eqref{eq:eps}
we obtain 
\begin{equation}
\label{eq:eos-hom}
\dot a=-\psi(a),
\end{equation} and then the singularity form in a finite amount of cosmological time if and only if \cite[Theorem 3.1]{Giambo:2005se}
\begin{equation}
\label{eq:sing-hom}
\int_0^{a_0} \frac{1}{\psi(a)}\,\mathrm da<+\infty
\end{equation}
for some $a_0>0$ corresponding to the initial data at $t=0$. 

Moreover, all the dynamical quantities involved can be expressed as a function of $a$, for example using \eqref{eq:G00}--\eqref{eq:G11} we have 
\begin{align}
\left(\frac{\mathrm d\phi}{\mathrm da}\right)^2&=\frac{2}{a^2}\left(1-\frac{a\,\psi'(a)}{\psi(a)}\right),\label{eq:psi-hom}\\
V(a)&=\frac{2\psi^2(a)}{a^2}\left(1+\frac{a\,\psi'(a)}{2\psi(a)}\right),\label{eq:V-hom}
\end{align}
whereas the mass in \eqref{eq:mass} takes the form
\begin{equation}
\label{eq:mass-hom}
m=\frac{r^3}2 a(t)\dot a^2(t)=\frac{r^3}2 a\,\psi(a)^2,
\end{equation}
which allows to write the trapped region $R<2m$ as 
\begin{equation}
\label{eq:trapp-hom}
\psi(a)^2 r^2>1,
\end{equation} 
and in this way one can find out sufficient conditions on the function $\psi(a)$ to develop a horizon or a naked singularity \cite[Theorem 5.2]{Giambo:2005se}.

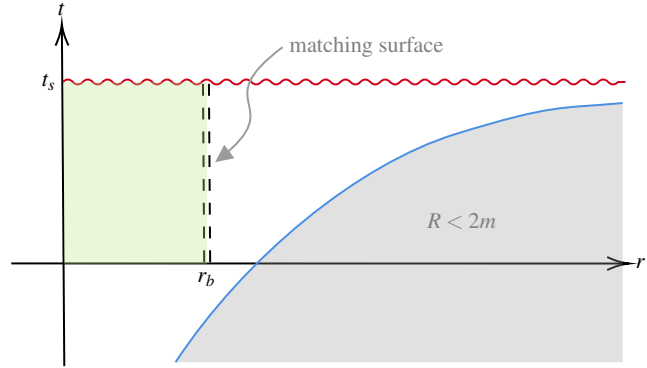
\begin{figure}
\centering

\tikzset{every picture/.style={line width=0.75pt}} 

\begin{tikzpicture}[x=0.75pt,y=0.75pt,yscale=-1,xscale=1]

\draw    (34.66,211.14) -- (344.12,211.2) ;
\draw [shift={(346.12,211.2)}, rotate = 180.01] [color={rgb, 255:red, 0; green, 0; blue, 0 }  ][line width=0.75]    (10.93,-3.29) .. controls (6.95,-1.4) and (3.31,-0.3) .. (0,0) .. controls (3.31,0.3) and (6.95,1.4) .. (10.93,3.29)   ;
\draw    (61.4,263) -- (60.18,92.25) ;
\draw [shift={(60.17,90.25)}, rotate = 89.59] [color={rgb, 255:red, 0; green, 0; blue, 0 }  ][line width=0.75]    (10.93,-3.29) .. controls (6.95,-1.4) and (3.31,-0.3) .. (0,0) .. controls (3.31,0.3) and (6.95,1.4) .. (10.93,3.29)   ;
\draw [color={rgb, 255:red, 208; green, 2; blue, 27 }  ,draw opacity=1 ]   (60.72,119.56) .. controls (62.39,117.89) and (64.05,117.89) .. (65.72,119.56) .. controls (67.39,121.23) and (69.05,121.23) .. (70.72,119.56) .. controls (72.39,117.89) and (74.05,117.89) .. (75.72,119.56) .. controls (77.39,121.23) and (79.05,121.23) .. (80.72,119.56) .. controls (82.39,117.89) and (84.05,117.89) .. (85.72,119.56) .. controls (87.39,121.23) and (89.05,121.23) .. (90.72,119.56) .. controls (92.39,117.89) and (94.05,117.89) .. (95.72,119.56) .. controls (97.39,121.23) and (99.05,121.23) .. (100.72,119.56) .. controls (102.39,117.89) and (104.05,117.89) .. (105.72,119.56) .. controls (107.39,121.23) and (109.05,121.23) .. (110.72,119.56) .. controls (112.39,117.89) and (114.05,117.89) .. (115.72,119.56) .. controls (117.39,121.23) and (119.05,121.23) .. (120.72,119.56) .. controls (122.39,117.89) and (124.05,117.89) .. (125.72,119.56) .. controls (127.39,121.23) and (129.05,121.23) .. (130.72,119.56) .. controls (132.39,117.89) and (134.05,117.89) .. (135.72,119.56) .. controls (137.39,121.23) and (139.05,121.23) .. (140.72,119.56) .. controls (142.39,117.89) and (144.05,117.89) .. (145.72,119.56) .. controls (147.39,121.23) and (149.05,121.23) .. (150.72,119.56) .. controls (152.39,117.89) and (154.05,117.89) .. (155.72,119.56) .. controls (157.39,121.23) and (159.05,121.23) .. (160.72,119.56) .. controls (162.39,117.89) and (164.05,117.89) .. (165.72,119.56) .. controls (167.39,121.23) and (169.05,121.23) .. (170.72,119.56) .. controls (172.39,117.89) and (174.05,117.89) .. (175.72,119.56) .. controls (177.39,121.23) and (179.05,121.23) .. (180.72,119.56) .. controls (182.39,117.89) and (184.05,117.89) .. (185.72,119.56) .. controls (187.39,121.23) and (189.05,121.23) .. (190.72,119.56) .. controls (192.39,117.89) and (194.05,117.89) .. (195.72,119.56) .. controls (197.39,121.23) and (199.05,121.23) .. (200.72,119.56) .. controls (202.39,117.89) and (204.05,117.89) .. (205.72,119.56) .. controls (207.39,121.23) and (209.05,121.23) .. (210.72,119.56) .. controls (212.39,117.89) and (214.05,117.89) .. (215.72,119.56) .. controls (217.39,121.23) and (219.05,121.23) .. (220.72,119.56) .. controls (222.39,117.89) and (224.05,117.89) .. (225.72,119.56) .. controls (227.39,121.23) and (229.05,121.23) .. (230.72,119.56) .. controls (232.39,117.89) and (234.05,117.89) .. (235.72,119.56) .. controls (237.39,121.23) and (239.05,121.23) .. (240.72,119.56) .. controls (242.39,117.89) and (244.05,117.89) .. (245.72,119.56) .. controls (247.39,121.23) and (249.05,121.23) .. (250.72,119.56) .. controls (252.39,117.89) and (254.05,117.89) .. (255.72,119.56) .. controls (257.39,121.23) and (259.05,121.23) .. (260.72,119.56) .. controls (262.39,117.89) and (264.05,117.89) .. (265.72,119.56) .. controls (267.39,121.23) and (269.05,121.23) .. (270.72,119.56) .. controls (272.39,117.89) and (274.05,117.89) .. (275.72,119.56) .. controls (277.39,121.23) and (279.05,121.23) .. (280.72,119.56) .. controls (282.39,117.89) and (284.05,117.89) .. (285.72,119.56) .. controls (287.39,121.23) and (289.05,121.23) .. (290.72,119.56) .. controls (292.39,117.89) and (294.05,117.89) .. (295.72,119.56) .. controls (297.39,121.23) and (299.05,121.23) .. (300.72,119.56) .. controls (302.39,117.89) and (304.05,117.89) .. (305.72,119.56) .. controls (307.39,121.23) and (309.05,121.23) .. (310.72,119.56) .. controls (312.39,117.89) and (314.05,117.89) .. (315.72,119.56) .. controls (317.39,121.23) and (319.05,121.23) .. (320.72,119.56) .. controls (322.39,117.89) and (324.05,117.89) .. (325.72,119.56) .. controls (327.39,121.23) and (329.05,121.23) .. (330.72,119.56) .. controls (332.39,117.89) and (334.05,117.89) .. (335.72,119.56) .. controls (337.39,121.23) and (339.05,121.23) .. (340.72,119.56) -- (344.38,119.56) -- (344.38,119.56) ;
\draw [color={rgb, 255:red, 74; green, 144; blue, 226 }  ,draw opacity=1 ][fill={rgb, 255:red, 155; green, 155; blue, 155 }  ,fill opacity=0.3 ]   (117.5,261) .. controls (152.15,208.32) and (209.5,161) .. (258.86,145.8) .. controls (308.22,130.59) and (311.56,133.01) .. (343.03,130.19) ;
\draw  [draw opacity=0][fill={rgb, 255:red, 155; green, 155; blue, 155 }  ,fill opacity=0.3 ] (117.5,261) -- (343.03,130.19) -- (343.03,261) -- cycle ;
\draw  [dash pattern={on 4.5pt off 4.5pt}]  (134.34,120.51) -- (134.99,214.48)(131.34,120.53) -- (131.99,214.5) ;
\draw  [draw opacity=0][fill={rgb, 255:red, 184; green, 233; blue, 134 }  ,fill opacity=0.3 ] (61.5,119.56) -- (133.16,119.56) -- (133.16,211.29) -- (61.5,211.29) -- cycle ;
\draw [color={rgb, 255:red, 155; green, 155; blue, 155 }  ,draw opacity=1 ]   (140.34,157.81) .. controls (175.25,130.08) and (132.5,131.25) .. (171.5,102) ;
\draw [shift={(137.5,160)}, rotate = 323.13] [fill={rgb, 255:red, 155; green, 155; blue, 155 }  ,fill opacity=1 ][line width=0.08]  [draw opacity=0] (8.93,-4.29) -- (0,0) -- (8.93,4.29) -- cycle    ;

\draw (243,184) node [anchor=north west][inner sep=0.75pt]  [color={rgb, 255:red, 128; green, 128; blue, 128 }  ,opacity=1 ] [align=left] {$\displaystyle R< 2m$};
\draw (58.72,119.56) node [anchor=east] [inner sep=0.75pt]  [font=\small] [align=left] {$\displaystyle t_{s}$};
\draw (133.16,214.29) node [anchor=north] [inner sep=0.75pt]  [font=\small] [align=left] {$\displaystyle r_{b}$};
\draw (173.5,102) node [anchor=west] [inner sep=0.75pt]  [font=\footnotesize,color={rgb, 255:red, 128; green, 128; blue, 128 }  ,opacity=1 ] [align=left] {matching surface};
\draw (348.12,211.2) node [anchor=west] [inner sep=0.75pt]  [font=\small] [align=left] {$\displaystyle r$};
\draw (60.17,87.25) node [anchor=south] [inner sep=0.75pt]  [font=\small] [align=left] {$\displaystyle t$};

\end{tikzpicture}

\caption{Homogeneous scalar field with a naked singularity. The region nearby the singularity $t=t_s$ is such that $R>2m$. In this way one can consider a region (in green) in a right neighbourhood $[0,r_b]$ of $r=0$ for the scalar field solution  and use Israel--Darmois condition to match this region with an exterior spacetime.}
\label{fig:homo-ns}
\end{figure}

The idea sketched above extends the approach used by Rituparno Goswami and Pankaj S. Joshi \cite{Goswami:2007na}, where $\rho_{SF}=\rho_0 \,a^{-\nu}$ with $\nu>0$ is considered. In the language of \eqref{eq:eps} that amounts to postulate
$$
\psi(a)= \psi_0\, a^{1-\nu/2},
$$
and in view of \eqref{eq:sing-hom} a singularity always forms in a finite amount of time if $\nu>0$. On the other side, if the inequality 
\begin{equation}
\label{eq:example-hom}
r<\frac{1}{\psi_0}a^{\tfrac{\nu}2-1} 
\end{equation}
is satisfied until $a=0$, then \eqref{eq:trapp-hom} cannot hold and then an apparent horizon with a trapped region cannot form. Then, if $\nu<2$, the right-hand side of \eqref{eq:example-hom} diverges in the approach to the singularity, and then $\forall r>0$ there is a region near the singular boundary that is untrapped, see Figure \ref{fig:homo-ns}. 

To build a global model one can consider a RW metric for $r\le r_b$ for some $r_b$, and get initial data $a(t)=a_0$ such that 
$$r_b<\tfrac{1}{\psi_0} a_0^{\tfrac\nu{2}-1}.$$
In this way the solution will live in the untrapped region $\forall r\in[0,r_b]$,  $\forall t\,:\,a(t)\in]0,a_0]$. Then a matching at $r=r_b$ using Israel--Darmois junction conditions can be performed with an exterior spacetime. This is made in  \cite{Goswami:2005fu} where generalized Vaidya spacetimes are considered, while in
\cite{Giambo:2005se} an anisotropic generalization of deSitter -- that is observed to be a subclass of the former -- is used.
The choice of the external solution is important also because it allows to evaluate the \textit{strength} of the singularity. Many definitions in this sense exists in literature, and we refer to \cite[Section 5.1]{Giambo:2005se} for a discussion about it, where a sufficient condition for the singularity  discussed here to be strong is given. As a consequence, in the example above sketched, there exists timelike geodesics terminating into a strong naked singularity. 

\subsection{Instability of the naked singularity}
\label{sec:naked}

The example discussed above has been further  developed in \cite{Goswami:2005fu}  where is shown that loop quantum gravity effects prevent the  formation of a singularity. 
But yet  at a classical level, the fact that they develop a naked singularity for  values of the parameter $\nu$ in an open interval of the real line, and therefore are stable with respect to that parameter's perturbation, could in principle be interpreted as a non-generic violation of cosmic censorship for these models. In the more recent \cite{Baier:2014ita} the problem is cast into a classical setting but adding a cosmological constant to \eqref{eq:EFE}, and finding once again naked singularities for a non-generic choice of a free parameter.

However, as already remarked, the genericity question crucially relies  on what we set as fixed and what we consider as ``variable''. For instance, in the above example, the free parameter is $\nu$, expressing the dependence of the energy density $\rho_{SF}$ from the scale factor $a$. But this does not keep the potential $V(\phi)$ fixed.
Indeed notice that, from \eqref{eq:psi-hom}--\eqref{eq:V-hom} and knowing the profile $\psi(a)$ one can find that $\phi$ diverges like $-\sqrt{\nu}\log a$ and then reconstruct back the original potential $V(\phi)$, that is given by
$$
V(\phi)=V_0\left(3-\frac{\nu}2\right)e^{\sqrt\nu \phi}.
$$
Things radically change if we \textit{fix} $V(\phi)$ and study the end state of homogeneous collapsing scalar field as a function of the initial data for $t=0$. 

Let us examine again the flat case $k=0$ and start by observing that  the choice of $(\phi_0,v_0)=(\phi(0),\dot\phi(0))$  fully determines the data at $t=0$, because in system \eqref{eq:dphi}--\eqref{eq:dh} the initial condition $h_0=h(0)$ is determined by the constraint $W(\phi_0,v_0,h_0)=0$ and the fact that we consider a collapse scenario. 
In the paper \cite{Giambo:2008ya} a class of $C^2$ potentials $V(\phi)$ is considered, that includes quite general polynomial potentials such as those asymptotically diverging like $\phi^{2n}$ as a subclass. In particular, $V(\phi)$ is supposed to satisfy (see \cite[Section II]{Giambo:2008ya}) the following properties: (i) it must be bounded from below, (ii) its critical points must be isolated and possible local maxima must be non-degenerate, (iii) a suitable compact sublevel set 
$\{V\le V_*\}$ must exist  and (iv) a growth condition on the first and second derivative of $\log V(\phi)$ at $\pm\infty$ is assumed to hold. 
Other class of potentials were considered in 
\cite{Giambo:2009zza}. 

Under the above assumptions is proved \cite[Theorem 3.7]{Giambo:2008ya} that -- except at most for an exceptional zero--measured set of initial data $(\phi_0,v_0)$ --  the scalar field diverges together with its derivative and develops a singularity at a finite amount of cosmological time. Moreover \cite[Theorem 4.1]{Giambo:2008ya} the scale factor $a(t)$ vanishes in the approach to the singularity but generically $\dot a(t)\to -\infty$. This means, recalling \eqref{eq:eos-hom}, that  \eqref{eq:trapp-hom} is always satisfied in a right neighbourhood of $a=0$, $\forall r>0$, and then an horizon is developing before the singularity. We stress that this happens up to a non-generic choice of initial data, once $V(\phi)$ is fixed.

The proof of the above results is carried on by observing first that, introduced the function
$$
y:=\frac{V(\phi)}{\tfrac12\dot{\phi}^2},
$$
then
 the scalar field energy $\rho_{SF}$ and the scale factor $a(t)$ satisfy the differential relations
\begin{subequations}
\begin{align}
\dot{\rho}_{SF}(t)&=2\sqrt 3\rho_{SF}^{3/2}(t)\frac1{1+y(t)},\label{eq:rho-sf}\\
\ddot a(t)&=-\frac{\dot a^2(t)}{a(t)}\left(\frac{2-y(t)}{1+y(t)}\right)
\label{eq:ddota}
\end{align}
\end{subequations}
Now the proof determines the asymptotic behaviour of $y(t)$, as follows:
\begin{enumerate}
\item first, it is shown that the scalar field $\phi(t)$ with collapsing initial data diverges and its derivative $\dot{\phi}(t)$ is eventually non-zero; then we can express -- at least eventually the dynamical quantities with respect to $\phi$;
\item the function $y$, seen as a function of $\phi$, satisfies a differential relation 
that is studied using the growth assumption on $V(\phi)$, finding that, up a non-generic choice of initial data, $y(\phi)\xrightarrow{\phi\to\infty} 0$;
\item using steps 1. and 2. in \eqref{eq:rho-sf}--\eqref{eq:ddota}, one finds that, for these non-generic data, $\rho_{SF}(t)$ diverges in a finite amount of cosmological time, and then a singularity develops, but $\dot a(t)$ is unbounded, which as observed above results generically in the formation of the horizon.
\end{enumerate}
The function $y(t)$ is basically a modified way to ``normalize'' the dynamical system, that is a classical approach to the system under suitable condition on the potential $V(\phi)$, see 
\cite{Foster:1998sk} where, 
instead of considering the unknown functions and \eqref{eq:dphi}--\eqref{eq:dh}, takes their ratios  to the Hubble function $h$, that in the flat $k=0$ case is connected, by \eqref{eq:W}, to the energy density $\rho_{SF}$.

It follows that, the cases depicted in \cite{Goswami:2007na} leading to a naked singularity, necessarily correspond to the non-generic case excluded by the analysis of step 2 above, see  \cite[Section V.]{Giambo:2008ya}). Same argument applies to the examples shown in 
\cite{Mosani:2021czj}. In this case the potential considered is $V(\phi)=\lambda\phi^2(\phi^2-\tfrac83)$, and it is prescribed $\rho_{SF}(a)=64\lambda(k-\log a)^2$ for $k>0$. Using the above relations it can be actually proved that the spacetime is regular for all $t>0$ -- although in \cite{Mosani:2021czj} it is observed that an ultra high density region is reached in a finite amount of time. However, once again, it can be seen that this situation   corresponds to the non-generic choice of initial data excluded by step 2 described above.
In fact, as observed in \cite[Example 3.9]{Giambo:2008ya}, if a potential $V(\phi)$ in this class is fixed, then initial data generically lead to black hole formation, and prescribing the dependence of $\rho(a)$ as done in \cite{Mosani:2021czj} precisely corresponds to the non-generic data  $(\phi(0),\dot\phi(0))$ that  possibly may not lead to the formation of an  horizon.

In conclusion, once a potential $V(\phi)$ obeying suitable -- yet rather inclusive -- conditions is prescribed, the collapse may end into a naked singularity only for a non-generic choice of the initial data on the scalar field. Nevertheless, the non-generic naked singularity can be gravitationally strong, which interestingly enough is the same conclusion reached in \cite{Guo:2020ked} for the critical collapse described before in Subsection \ref{sec:crit-ns}. 

\section{Other developments and conclusions}\label{sec:stuff}
The literature discussed in this review, although covering some of the undoubtedly recognized as cornerstones in the analytical and numerical studies on scalar field collapse, is necessarily a selection in the huge amount of contributions on this subject.

Concerning the analytical studies, we have already cited some recent developments aiming to prove stability of spherically results with respect to general -- hence possibly nonspherical -- perturbations \cite{Kilgore:2021loy,Luk2018,Luk2019-I} in the free massless case.
 Just to name a few other analytical results, when a potential is added, collapse in cylindrical symmetry under self--similarity assumptions has been discussed in \cite{Condron:2013yaa,Condron:2013hja,Condron:2013rha}. With aim to study the big bang singularity, backwards evolution of a free massless scalar field is considered in \cite{Rodnianski:2014yaa}, where the initial data are close to the homogeneous case and given on a 3-dimensional torus $\mathbb T^3$. In \cite{Zhang:2015rsa}, a scalar field self--interacting with a negative exponential potential is Kaluza--Klein lifted to a vacuum solution in $\mathbb R^{1+3}\times S^2$, to find singular solutions without trapped surface formation developing from regular data. Notice that the family of 4-dimensional scalar field solutions considered contains an example that extends to a case with potential the self--similar naked singularity of \cite{Christodoulou:1994hg}. The stability of this important example, as we have said, remains a problem with some issues to be fully understood, although \cite{Chr99a} represent of course a crucial step. When confining to the self--similar collapse, the situation has been studied e.g. in \cite{Nolan:2006my,Banerjee:2017njk}.
 It is worth citing also  \cite{Li:2017rgs} that attempts a first important generalization of the naked singularity instability  result when no symmetry on the spacetime geometry is assumed. 

The PDE approach that has been presented here is not the only one used, even in an analytical context. Some results based on Fuchsian techniques can be found in \cite{Andersson:2000ay,Damour:2002tc}, while an interesting approach involving inverse reconstruction methods recently appeared \cite{Kurylev:2022mcq} where no particular geometric symmetries are required. Also numerical studies have tried to cover solutions not necessarily spherically symmetric, in the attempt to reproduce the critical behaviour also in more general cases, see e.g. \cite{Healy:2013xia}.

As already said, the study of scalar field cosmologies via dynamical system techniques also are a huge part of this subject, although many are understandably focused on expanding solutions. Having recalled once again reviews \cite{Leon:2020pfy,Leon:2020ovw} for extensive treatises in this context, let us cite the paper \cite{Hernandez:2018rxc} where back--reaction terms are considered to find collapsing solutions that do not develop singularities. 
A great deal of attention has been also paid to (re)collapsing models where a matter contribution, typically a perfect fluid -- either coupled or uncoupled to the scalar field -- is added, mostly arising from higher order gravity theory \cite{Miritzis:2003eu,Miritzis:2005hg}. The problem has been studied in \cite{Tavakoli:2013dwz} for the case of positive exponential potentials and in \cite{Giambo:2008sa} for the same class of potentials studied in \cite{Giambo:2008ya}, where again black hole formation is shown to generically form. As one may expect, big crunch singularities naturally arise also when one deals with negative potentials \cite{Giambo:2014jfa,Giambo:2015tja}. Coupling effect may be mimicked also using deformation of the phase space, producing corrections  to the dynamical system also in absence of a perfect fluid, as studied in \cite{Rasouli:2013sda}, where anyway the deformation is shown to prevent, in some cases, the formation of singularities from gravitational collapse.  

There are other approached to the problem worth to be cited, for instance  \cite{Bhattacharya:2009bz} where the scalar field is assumed to be timelike, in such a way that comoving coordinates can be built considering a possibly inhomogeneous spacetime, although $\phi=\phi(t)$ only.

In summary, the collapse of a scalar field has been a valuable tool for investigating various aspects of gravitational physics. The scalar field's multiplicity and rich physical interpretations, both alone and when coupled with other matter models, have contributed to its continued relevance. Moreover, the associated mathematical model is relatively simple, particularly in the context of spherical symmetry, making it a useful test-bed for exploring classical and nonlinear  gravity theories. As such, the collapse of a scalar field has been and continues to be an important means of probing physics in the theory of gravitation.



\end{document}